\documentclass{aa}
\usepackage[dvips]{graphics}

\begin{document}

   \title{The elementary spike produced by a pure $e^+e^-$ pair-electromagnetic pulse from a Black Hole: The PEM Pulse}


   \author{Carlo Luciano Bianco\thanks{e-mail: bianco@icra.it}
          \inst{1}
          \and
          Remo Ruffini\thanks{e-mail: ruffini@icra.it}
          \inst{1}
	        \and          
	        She-Sheng Xue\thanks{e-mail: xue@icra.it}
          \inst{1}
          }

   \offprints{R. Ruffini}
   \institute{I.C.R.A.-International Center for Relativistic Astrophysics
and
Physics Department, University of Rome ``La Sapienza", I-00185 Rome,
Italy
             }

\authorrunning{C.L. Bianco \and R.Ruffini \and S.-S. Xue}

\titlerunning{Elementary spike from an EMBH Black Hole: the PEM pulse}

  \date{Received (to be inserted later) /Accepted (to be inserted later)}

\abstract{In the framework of the model that uses black holes endowed with electromagnetic structure (EMBH) as the energy source, we study how an elementary spike appears to the detectors. We consider the simplest possible case of a pulse produced by a pure $e^+e^-$ pair-electro-magnetic plasma, the PEM pulse, in the absence of any baryonic matter. The resulting time profiles show a {\em Fast-Rise-Exponential-Decay} shape, followed by a power-law tail. This is obtained without any special fitting procedure, but only by fixing the energetics of the process taking place in a given EMBH of selected mass, varying in the range from 10 to $10^3$ $M_\odot$ and considering the relativistic effects to be expected in an electron-positron plasma gradually reaching transparency. Special attention is given to the contributions from all regimes with Lorentz  $\gamma$ factor varying from $\gamma=1$ to $\gamma=10^4$ in a few hundreds of the PEM pulse travel time. Although the main goal of this paper is to obtain the elementary spike intensity as a function of the arrival time, and its observed duration, some qualitative considerations are also presented regarding the expected spectrum and on its departure from the thermal one. The results of this paper will be comparable, when data will become available, with a subfamily of particularly short GRBs not followed by any afterglow. They can also be propedeutical to the study of longer bursts in presence of baryonic matter currently observed in GRBs.
}

   \maketitle

\section{Introduction}\label{int}

It is by now clear that the Gamma-Ray Bursts phenomena, here generally called GRBs, far from representing a single physical event, are quite complex and are composite of very different phases, corresponding to distinctively different physical processes. Any data analysis procedure should be done taking into due account these different epochs and avoiding time averaging procedures on different epochs.

In recent years much attention has been devoted to analyzing GRB afterglows, originating from the interaction of a relativistic expanding fireball with the surrounding baryonic material, leading to important correspondences between observational data and phenomenological models (see e.g. \cite{V98}, \cite{SE00} and \cite{FDmg9}).

Less progress has been made in understanding the basic mechanism which produces the fireball and in explaining the structure of the burst itself. In the existing literature, it has been alternatively assumed that GRBs originate:\\
{\bf a)} by the smooth transition to transparency of an optically thick electron-positron plasma, whose origin is not discussed (see e.g. \cite{g86});\\
{\bf b)} from external shock processes in the collision of an expanding fireball, whose origin is not discussed, with baryonic matter at rest (see e.g. \cite{Mr92});\\
{\bf c)} from the interaction with each other of different parts of the fireball due to different expansion velocities, by internal shock processes (see e.g. \cite{Px94});\\
{\bf d)} from a mixed ``internal-external'' scenario, in which both effects are present (see e.g. \cite{p99});\\
{\bf e)} from a Compton-drag process, caused by the interaction of a relativistic fireball, whose origin is not discussed, with a very dense soft photon bath (see e.g. \cite{glcr00}).

Important as they are, these semi-empirical approaches miss the completeness of a comprehensive quantitative model. Some of the above processes could indeed be present in the different epochs of the GRBs. In order to evaluate their relative magnitude and give a quantitative estimate of the associated relativistic effects is necessary to have a detailed model, with the corresponding equations of motion and the time evolution of the system.

For this reason we consider the model which assumes the energy source of GRBs to be the process of vacuum polarization around a black hole endowed with electromagnetic structure (EMBH) (see \cite{dr75} and \cite{rukyoto}). For such a system the fundamental energetic aspects have been clarified by the introduction of the Dyadosphere of an EMBH (\cite{prxb}) as well as the equations of motion for the system have been integrated and are summarized in the next section. 

We proceed in four separate steps:

In this paper, we make a first step. We consider the simplest possible case: the one corresponding to a pulse produced by a pure $e^+e^-$ pair-electro-magnetic plasma, the PEM pulse, in the absence of any baryonic matter.

The equations of motion of this system have already been integrated (\cite{rswx99}). We here examine how the PEM pulse gradually emits radiation during its evolution and the GRB simply occurs during the smooth approach to the transparency condition, when the mean free path of the photons in the plasma, $L_\gamma$ in the comoving frame and $\lambda$ in the laboratory frame, given by
\begin{equation}
L_\gamma=\frac{1}{\sigma_{\gamma,e}n_{e^\pm}}, \; \; \; \; \lambda=\frac{L_\gamma}{\gamma},
\label{Lgamma}
\end{equation}
approaches the thickness of the PEM pulse itself. Here $\sigma_{\gamma,e}$ is the photon and electron Compton scattering cross-section, whose Thomson limit is $\frac{\pi\alpha^2}{m^2}\simeq 6.66\cdot 10^{-25}\, {\mathrm cm^2}$ and $n_{e^\pm}$ is the proper number density of $e^+e^-$ pairs. Thus the GRB is essentially determined by the electron-positron pair annihilation and the expansion and cooling of the PEM pulse.

It has been clear since the classic work of Rees on expanding radio sources that relativistic effects are central to the understanding of the astrophysics of extragalactic sources (\cite{r66}). The typical Lorentz  $\gamma$ factor considered in that study was of the order of $\gamma \sim 5$ and essentially constant over the time of observation of the astrophysical source. In the case of GRBs such relativistic effects are also important, but there are three major differences:
\begin{enumerate}
\item The relativistic effects are much more extreme. It was pointed out (see \cite{rswx99}) that, in the case of the absence of baryonic matter, a Lorentz $\gamma$ factor of up to $10^4$ can be reached. This result has also been extended to the case of the expansion of the PEM pulse in the presence of baryonic matter (see \cite{rswx00}) where even larger values of the Lorentz $\gamma$ factor can be reached.
\item The transition from $\gamma = 1$ to $\gamma \sim 10^4$ occurs in a few hundreds of the characteristic travel times in the PEM pulse (i.e. a few seconds). It is therefore impossible, as shown in Fig. \ref{ETSNCF}, in Fig. \ref{ETSCONF} and in sections \ref{arrival}, \ref{arrivalf}, \ref{arrivalft}, \ref{t90} of the present paper, even as a rough approximation, to consider the Lorentz $\gamma$ factor to be constant during the emission process of the GRB.
\item The transparency condition given by Eq.(\ref{Lgamma}) is reached gradually by the effect of the cooling and the expansion of the plasma and the corresponding decrease in the number of the electron-positron pairs. Consequently, in principle, all the stages from $\gamma = 1$ to $\gamma \sim 10^4$ contribute to the final burst structure.
\end{enumerate} 
 
The main point of the present paper is to clarify the interplay of relativistic effects at work in this extreme case, in order to obtain the elementary spike intensity as a function of the arrival time, and its observed duration.  In particular we here take into account, still for simplicity in the case of a spherical geometry, the effects due to:
\begin{itemize}
\item the varying thickness of the emitting region,
\item the energy flux, essentially modulated by a time varying ``screening factor'' (see appendix A and B),
\item the time variation  of the $\gamma$ factor and consequences for the observed arrival time.
\end{itemize}
We examine explicitly, for simplicity, the case of Reissner-Nordstr\"{o}m EMBHs with $\frac{Q}{M}=0.1$ and $M=10M_\odot$, $M=10^2M_\odot$ and $M=10^3M_\odot$. Some considerations on the expected spectrum and its departure from the thermal one are also presented in section (\ref{spectraG}) and will be further examined in forthcoming publications.

All the results of the treatment presented in this paper can be observationally relevant for a very special class of short GRBs without any afterglow. They can also be of qualitative interest for the first spike and early features of a more complex long burst. It is important to stress that in the case of the elementary spike here considered the entire energy of the Dyadosphere is emitted in the burst. This is not the case for the long bursts, the large majority of the currently observed GRBs, where much of such energy is transferred to the kinetic energy of the baryonic component.

In a forthcoming paper (Bianco, Ruffini, Xue, in preparation), we consider the emission from a PEM pulse interacting with baryonic matter, before the condition of transparency be reached, and the corresponding consequences for the observable effects on the intensity, spectrum and time structure of the burst. The equations of motion for such a pair-electromagnetic-baryonic pulse (PEMB Pulse) have been integrated and the relative intensity to be expected for the GRB versus the kinetic energy left in the accelerated baryonic material (\cite{rswx00}) has also been given.

In a final paper (Bianco, Chardonnet, Fraschetti, Ruffini, Xue, in preparation) we analyze the interaction of the accelerated baryonic material (ABM Pulse) with the interstellar medium. As the transparency condition is reached all the baryonic matter is left having acquired an enormous Lorentz $\gamma$ factor reached at the time of decoupling, typically in the range $10 \sim 10^4$. The interaction of these very high energy baryons with the surrounding interstellar medium gives rise to the afterglow.

We finally consider (Chardonnet \& Ruffini, in preparation), within the above model, the production of very high energy cosmic rays by the electrostatic acceleration process of the remnant EMBH.

\section{Main assumptions of the EMBH model}\label{maembhm}
 
The most general black hole expected from gravitational collapse is an EMBH characterized by a Kerr-Newmann geometry endowed with axial symmetry and electric and magnetic fields (\cite{rw71}). That indeed an EMBH with mass smaller than $7.2\cdot 10^6 M_\odot$, can give rise to vacuum polarization and to the creation of $e^+e^-$ pair-electromagnetic plasma outside the horizon via the Heisenberg-Euler-Schwinger process was clearly demonstrated (see \cite{dr75}). It was there shown how these process can approach reversibility, in the sense of Christodoulou and Ruffini (\cite{cr71}) and that this phenomenon would lead to a most natural model for GRB.

The discovery of the afterglow of GRB by the Beppo-Sax satellite and the consequent clear determination of the distance and energetics of the GRB's has motivated us to return to this field. In order to give an estimate of the efficiency of the quantum electrodynamical process in the gravitational field of an EMBH we have considered the idealized case of a spherically symmetric Reissner-Nordstr\"{o}m geometry, neglecting all the effects associated with rotation, to be later examined in a much more complex treatment. We have then computed the physical parameters, the spatial extension, the total energy, and the spectrum of the relativistic plasma of electron-positron pairs created by the vacuum polarization process and introduced the concept of the Dyadosphere (\cite{prxa}).

The computation of the relativistic hydrodynamics of such an $e^+$ and $e^-$ pairs and electromagnetic radiation plasma in the field of an EMBH, still in the simplified case of spherical symmetry, have been carried out both with simple semi-analytical and numerical approach carried out in Rome and their validation by the full hydrodynamical numerical computation at Livermore (see \cite{rswx99}). The main result shows the formation of a  sharp slab of $e^+$ and $e^-$ pairs and electromagnetic radiation: the PEM pulse. Such a PEM pulse keeps a  constant width in the laboratory frame and in a few hundreds of characteristic crossing times of the Dyadosphere reaches extreme relativistic conditions with characteristic  Lorentz $\gamma$ factors of $10^3$ - $10^4$. The pair density at the beginning is so high that the plasma is optically thick, and only very few photons can escape from the plasma. With the expansion, cooling and annihilation of electron-positron pairs, the condition of transparency is gradually reached (see Eq.(\ref{Lgamma})). The details of the temporal development of the burst emitted as the transparency condition is reached, as well some indications on the expected spectra, are the subject of the present paper. For simplicity, we have neglected the feedback of the PEM pulse on the adiabaticity condition prior to reaching the final moment of transparency.

The case in which some baryonic matter is engulfed by the PEM pulse, prior to reaching the condition of transparency, has been the subject of a successive work (\cite{rswx00}). Our simplified model, as well as the validation by the full numerical codes at Livermore, have shown how the slab approximation is still valid in this more general case for a large range of the masses of the baryonic matter (a pairs-electromagnetic-baryons pulse, PEMB pulse). Most important, the addition of baryonic matter leads, through the electronic component, to an increase in the opacity of the PEMB pulse. The transparency condition is now reached at later times, leading to an ever increasing transfer of the total energy of the PEMB pulse to the kinetic energy of the baryonic component. The larger the amount of baryonic matter, the smaller is the energy of the PEMB pulse released in the GRB, for fixed values of the mass and charge of the EMBH.

All these treatments refer to the idealized case of an already formed EMBH, it is clear, however, that in reality the Dyadosphere will be formed during the process of gravitational collapse itself, prior to the formation of the EMBH, and such a process may have some distinct detailed observational signature. The general energetic features of GRBs here presented have been obtained from the idealized model which considers as a starting point an already formed black hole (\cite{prxb}). In order to obtain the fine details observed in the time structure of GRBs, as well as their detailed spectral evolution, there is no alternative but to study the gradual formation of the Dyadosphere, as the horizon of the EMBH is approached and formed. In preparation of this more complicated analysis and the one corresponding to axially symmetric configurations, we have obtained some preliminary results concerning the relevant physics of a charged collapsing shell in general relativity (Klippert and Ruffini, in preparation). The general energetic aspects here considered will not be modified but the time constant will be longer due to general relativistic effects (Cherubini, Jantzen, Ruffini, in preparation).

\section{The radiation flux from the PEM pulse}\label{flux}

The frequency $\omega$ and wave-vector ${\bf k}$ of photons emitted from the PEM pulse (see Fig. \ref{costr}) expressed in the laboratory frame are:

\begin{equation}
{\bf k} = \frac{\omega}{c} \left(-\sin\vartheta{\bf u} +\cos\vartheta{\bf v}\right), \quad \left|{\bf k}\right|=\frac{\omega}{c},
\label{kdirection}
\end{equation}
where $\vartheta$ is the angle (in the laboratory frame) between the radial expansion velocity and the direction from the origin of the PEM pulse to the observer, ${\bf v}$ is a unit vector along the radial expansion velocity of the PEM pulse, and ${\bf u}$ is a unit vector orthogonal to ${\bf v}$ oriented toward rising $\vartheta$. We are assuming here that ${\bf k}$ and $R_{\rm T}$ are parallel, also for photons emitted with $\vartheta\neq 0$. This is clearly a good approximation, because the distance $R_{\rm T}$ corresponds to a redshift $z\sim 1$, while the radius of the emitting region is of the order of magnitude of a few light seconds.

Then the Lorentz boost along ${\bf v}$ to the comoving frame of the PEM pulse yields the corresponding comoving quantities:

\begin{equation}
\omega _ \circ   = \gamma \omega \left( {1 - \frac{v}{c}\cos \vartheta } \right),\quad \omega _ \circ   = \left| {{\bf k}_\circ } \right|c,
\label{entran}
\end{equation}
\begin{equation}
{\bf k}_\circ = -\left |{\bf k}\right|\sin\vartheta{\bf u}+\gamma \left|{\bf k}\right|\left(\cos\vartheta-\frac{v}{c}\right){\bf v},
\label{kntran}
\end{equation}

In the comoving frame photons radiating out of the PEM pulse must have (see Eq.(\ref{kntran})):

\begin{equation}
\cos\vartheta\ge {v\over c},
\label{cos}
\end{equation}
because the component of the photon momentum in the comoving frame along the radial expansion velocity direction must be positive in order to escape.

The large amount of high-energy photon emission is mainly due to electron-positron annihilations. Thus, as a preliminary consideration and approximation, we assume that photons are in equilibrium at the same temperature $T$ with electron-positron pairs before and at decoupling. In the comoving frame of the photon electron-positron pair plasma fluid, the black-body spectrum of photons that are in thermal with $e^+e^-$-pairs is given by

\begin{equation}
\frac{dn_\gamma}{d^3k_\circ}= \frac{1}{\pi^2}\frac{1}{{\rm exp}\left(\frac{\hbar\omega_\circ}{kT}\right)-1},
\label{cspectrum}
\end{equation}
where $n_\gamma$ is the number-density of photons and $T$ is the temperature in the comoving frame. The non-thermal spectrum due to multiple inverse Compton scattering may slightly modify these assumptions. These consideration relevant to the details of the spectrum (see section \ref{spectraG}) do not modify the temporal profile of the total detected radiation flux, which largely depends on the screening factor $S(t)$ rather than on the radiative spectrum.

Note that due to the Liouville theorem on the invariance of the distribution function (see e.g. \cite{eh}), we can write:

\begin{equation}
\frac{1}{{\rm exp}\left(\frac{\hbar\omega_\circ}{kT}\right)-1}=\frac{1}{{\rm exp}\left(\frac{\hbar\omega}{kT_{\rm lab}}\right)-1}
\label{liouville}
\end{equation}
where $\omega_\circ$ and $T$ are the photon energy and temperature in the comoving frame, while $\omega$ and $T_{\rm lab}$ are the same quantities in the local laboratory frame. We can now use Eq.(\ref{entran}) for $\omega_\circ$ obtaining:

\begin{equation}
T_{\rm lab}=\frac{T}{\gamma \left(1-\frac{v}{c}\cos\vartheta\right)} .
\label{tspectrum}
\end{equation}

Therefore we can use the following formula for the photon spectrum:

\begin{equation}
\frac{dn_\gamma}{d^3k}= \frac{1}{\pi^2}\frac{1}{{\rm exp}\left(\frac{\hbar\omega\gamma \left(1-\frac{v}{c}\cos\vartheta\right)}{kT}\right)-1} .
\label{lspectrum}
\end{equation}

Integrating Eq.(\ref{lspectrum}) over all photon momenta, we obtain the total radiation flux in the local laboratory frame
of the PEM pulse.

The number of photons ($k\rightarrow k+dk$) radiating per unit time from a small surface element $d^2\Sigma$ of the PEM pulse to the detector is given by (per unit area of detector):

\begin{equation}
\frac{dn_\gamma}{d^3k}(\hat k\cdot {\bf v})cd^2\Sigma\frac{d\Omega_k}{A}, \quad d\Omega_k=\frac{A}{R_{\rm T}^2}
\label{too}
\end{equation}
where $\hat k=\frac{{\bf k}}{\left| {\bf k} \right|}$ is directed towards the observer, $A$ is the area of detector,
$R_{\rm T}$ the distance to the detector from the origin of the PEM pulse and $d\Omega_k$ is the solid angle subtended by the detector. From Eqs.(\ref{cos},\ref{kdirection}), we have

\begin{equation}
\hat k\cdot {\bf v}=\cos\vartheta\in \left[\frac{v}{c},1\right].
\label{kv}
\end{equation}

By using Eq.(\ref{lspectrum}), we integrate Eq.(\ref{too}) over all photon energies and obtain the observed infinitesimal energy flux in the case of an optically thin PEM pulse $\left({\rm in}\; \frac{\rm erg}{{\rm cm}^2{\rm s}}\right)$:

\begin{eqnarray}
dF^{\rm T} \left( {t,\vartheta } \right) = \frac{1}{{R_{\rm T}^2 }}a\left( {\frac{{T\left( t \right)}}{{\gamma \left( t \right)\left( {1 - \frac{v\left( t \right)}{c}\cos \vartheta } \right)}}} \right)^4 \cdot\nonumber\\
\cdot c \, \cos \vartheta \, d^2 \Sigma .
\label{bssll}
\end{eqnarray}

Integrating only $d\phi\in \left[0,2\pi\right]$ in the surface-element $d^2\Sigma$ of the PEM pulse, and adding the screening factor $S\left(t\right)$, we finally obtain the infinitesimal photon energy flux directed towards the detector and emitted at time $t$ with an angle $\vartheta$ for an optically thick PEM pulse:

\begin{eqnarray}
dF \left( {t,\vartheta } \right) = 2 \pi a\left( {\frac{{T\left( t \right)}}{{\gamma \left( t \right)\left( {1 - \frac{v\left( t \right)}{c}\cos \vartheta } \right)}}} \right)^4 \cdot\nonumber\\
\cdot r^2 \left( t \right)c\cos \vartheta S\left( t \right)d\cos \vartheta ,
\label{bssl}
\end{eqnarray}
where we omit the factor $R_{\rm T}^{-2}$. $dF$ will be then measured in terms of $\frac{{\rm erg}}{{\rm s}}$, and we will still have to multiply the result by $R_{\rm T}^{-2}$ to obtain the observed intensities. Note that the radiation flux is axially symmetric with respect to the axis directed towards the detector.
 
\section{The arrival time versus emission time}\label{arrival}

\begin{figure}
\begin{center}
\resizebox{\hsize}{!}{\includegraphics{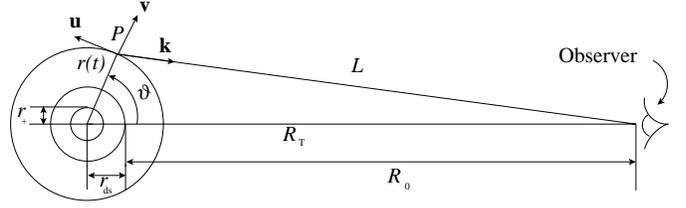}}
\caption{Scheme to find the relation between the emission time $t$ in the local laboratory frame and the arrival time $t_{\rm a}$. $P$ is the point where a generic photon is emitted. $L$ is the distance of $P$ from the observer. $R_{\rm T}$ is the distance of the black hole from the observer. $r_{\rm ds}$ is the radius of the dyadosphere. $r_{\rm +}$ is the radius of the horizon. $r\left( t \right)$ is the radius of the external surface of the fireball at time $t$. $R_0$ is defined by $R_0 \mathop  \equiv \limits^{def.} R_{\rm T} - r_{\rm ds}$. $\vartheta$ is the angle between $r\left( t \right)$ and $R_{\rm T}$. ${\bf v}$ is a unit vector along the radial expansion velocity. ${\bf u}$ is a unit vector orthogonal to ${\bf v}$ oriented toward rising $\vartheta$. ${\bf k}$ is the momentum of the photons emitted toward the observer. Note that we have assumed ${\bf k}\parallel R_{\rm T}$, and this is certainly a very good approximation (see text, section \ref{flux}).}
\label{costr}
\end{center}
\end{figure}

Due to the high value of the Lorentz $\gamma$ factor $\left(10^3 \sim 10^4\right)$ for the bulk motion of the expanding PEM pulse, the spherical waves emitted from its external surface appear extremely distorted to a distant observer (see e.g. \cite{r66}). The surface emitting the photons detected at an arrival time difference $t_{\rm a}$, measured from the arrival of the first photon, is not trivially the spherical surface of the fireball at a fixed time difference $t$, measured from the initial emission, but photons emitted at different $t$ and at different angles reach the detector at the same time $t_{\rm a}$. The relation between emission time $t$ and arrival time $t_{\rm a}$ in the case of a constant $\gamma$ for the expanding fireball, has been found by Rees (for the definition of $\vartheta$ see Fig. \ref{costr}) (see \cite{r66}):

\begin{equation}
t_{\rm a}  = t\left( {1 - \frac{v}{c}\cos \vartheta } \right) .
\label{tar}
\end{equation}

The external radius $r\left( t \right)$ of the fireball, due to the constancy of $\gamma$, is obviously given by:

\begin{equation}
r\left( t \right)=vt
\label{rvt}
\end{equation}

If, in Eq.(\ref{tar}), we fix the value of the arrival time $t_{\rm a}=t_{\rm a}^\star$, using Eq.(\ref{rvt}) we can find the equation describing the ``surface'' emitting the photons detected at arrival time $t_{\rm a}^\star$:

\begin{equation}
r= \frac{{v\;t_{\rm a}^\star }}{{1 - \frac{v}{c}\cos \vartheta }}  ,
\label{ETSC}
\end{equation}
which describes an ellipsoid of eccentricity $\frac{v}{c}$ (see \cite{r66}).

This is the approximation that has been widely used in the gamma ray burst literature: see e.g. \cite{Sa97}, \cite{Sua97}, \cite{RRa99}, \cite{Fa96}, \cite{Ga99}, \cite{F99}, \cite{Pa98}, \cite{p99} and references therein.

However, in our case the velocity $v$ is not constant: in a few hundreds of the PEM pulse crossing time\footnote{The crossing time of the PEM pulse is given by $\frac{r_{\rm ds}-r_{\rm +}}{c}$, and this is $\sim 10^{-2}$ seconds, for a black hole with $M=10^3M_\odot$ and $Q=0.1Q_{\rm max}$} the $\gamma$ factor of the bulk motion of the expanding PEM pulse goes from $0$ to $\sim 10^4$, and so we have to find the generalization of Eq.(\ref{tar}) for nonconstant velocity. This can be done using the geometry of Figure \ref{costr}.

We set $t=0$ when the fireball starts to expand and the first photons are emitted, so that $r\left( 0 \right)=r_{\rm ds}$.
Let a photon be emitted at time $t$ from the point $P$. Its distance from the observer is $L$. The time it takes to arrive at the detector is, of course, $\frac{L}{c}$. So, its arrival time, measured from the arrival of the first photon a time $\frac{R_0}{c}$ after its emission at $t=0$, is:
\begin{equation}
t_{\rm a}=t+\frac{L}{c}-\frac{R_0}{c}
\label{ta}
\end{equation}
where we have defined $t_{\rm a}=0$ when the first photon emitted at $t=0$ and $\vartheta=0$ reaches the observer. $L$ is clearly given by:
\begin{equation}
L = \sqrt {R_{\rm T}^2  + r\left( t \right)^2  - 2\,R_{\rm T}\,r\left( t \right)\cos \vartheta } 
\label{Lex}
\end{equation}
where at any given value of emission time $t$, $\cos \vartheta$ can assume any value between $\frac{v\left( t \right)}{c}$ and $1$ as noted above, where $v\left( t \right)$ is the expansion speed of the fireball at time $t$ (see Eq.(\ref{cos})). Now, $r\left( t \right)$ is of the order of magnitude of some light-seconds, and $R_{\rm T}$ corresponds to a redshift $z\sim 1$. So we can expand the right hand side of equation (\ref{Lex}) in powers of $\frac{r\left( t \right)}{R_{\rm T}}$ at first order:
\begin{equation}
L \simeq R_{\rm T}\left( {1 - \frac{{r\left( t \right)}}{R_{\rm T}}\,\cos \vartheta } \right),
\label{Lapp}
\end{equation}
which corresponds to neglecting the lateral displacement from the line of sight axis.

Substituting  (\ref{Lapp}) into (\ref{ta}) yields:
\begin{equation}
t_{\rm a}  = t +  \frac{{r_{\rm ds} }}{c} - \frac{{r\left( t \right)}}{c}\cos \vartheta ,
\label{taapp}
\end{equation}
where we have used the fact that $R_{\rm T}=R_0+r_{\rm ds}$ (see Figure \ref{costr}).
For $r\left( t \right)$ we can use the following expression:
\begin{equation}
r\left( t \right) = \int_0^t {v\left( {t'} \right)dt'}  + r_{\rm ds} ,
\label{rdiv}
\end{equation}
so that equation (\ref{taapp}) can be written in the form:
\begin{equation}
t_{\rm a}  = t - \frac{{\int_0^t {v\left( {t'} \right)dt'}  + r_{\rm ds} }}{c}\cos \vartheta  + \frac{{r_{\rm ds} }}{c},
\label{ta_fin}
\end{equation}
which reduces to Equation (\ref{tar}) only if $v$ is constant and $r_{\rm ds}$ is negligible with respect to $r\left( t \right)$.

\begin{figure}
\resizebox{\hsize}{!}{\includegraphics{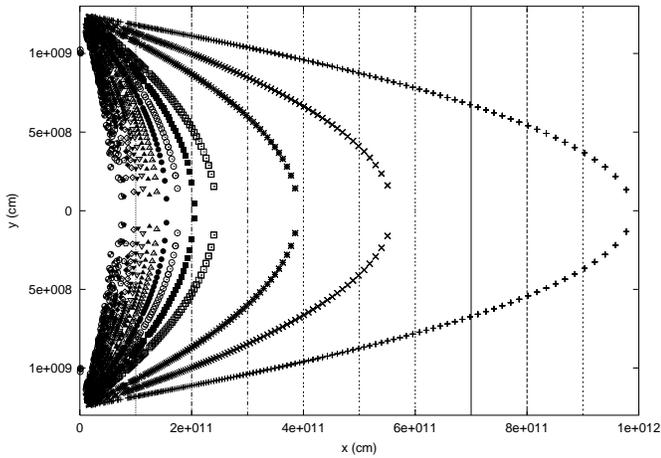}}
\caption{Equitemporal surfaces, projected into the plane of Fig. \ref{costr}, with Cartesian coordinates centered on the black hole, for a PEM pulse arising from a black hole with $M=10^3M_\odot$ and $Q=0.1Q_{\rm max}$, neglecting the fireball thickness effects (see Fig. \ref{1000_wrong} and par. \ref{arrivalf}), with $t_{\rm a}^\star$ going from 0.02620 seconds (the inner surface) to 0.02662 seconds (the outer one) with steps of $2\cdot 10^{-5}$ seconds. The observer is far away along the $x$ axis, while $y$ axis scale is expanded to see the extremely elongated surfaces, making the spherical surfaces centered on the black hole appear to be the vertical lines given in the figure.}
\label{ETSNCF}
\end{figure}

Also in Eq.(\ref{ta_fin}) we can fix $t_{\rm a}=t_{\rm a}^\star$ to obtain the equation describing the surface that emits the photons detected at an arrival time $t_{\rm a}^\star$. In this case, we no longer have ellipsoids of constant eccentricity $\frac{v}{c}$. Since the velocity is strongly varying from point to point, we have more complicated surfaces like the profiles reported in Fig. \ref{ETSNCF} where at every point there will be a tangent ellipsoid of constant eccentricity, but such an ellipsoid varies in eccentricity from point to point.

For a fixed time $t$ of emission in Eq.(\ref{ta_fin}), the allowed angular interval $\frac{v}{c} \le \cos\vartheta \le 1$ leads to a corresponding smearing of the arrival time $t_a$ over the interval
\begin{equation}
\Delta t_a = \frac{r}{\gamma^2c\left(1+\frac{v}{c}\right)},
\label{angular}
\end{equation}
called the {\em angular time scale} (see e.g. \cite{p99}).

\section{The radiation flux with respect to arrival time for an infinitely thin fireball}\label{arrivalf}

Eq.(\ref{bssl}), integrated with respect to $\vartheta$, gives us the value of $F \left( t \right)$,  the flux emitted from the PEM pulse at time $t$. Instead we want to compute  $F \left( t_{\rm a} \right)$,  the flux detected at an arrival time $t_{\rm a}$. In principle, we should substitute Eq.(\ref{ta_fin}) into Eq.(\ref{bssl}) before the integration, but we have no analytical relation between $v \left( t \right)$ and $t$, so a numerical integration must be performed. We used the following approach:
\begin{enumerate}
\item We fix the value of the emission time $t$, and let $\cos \vartheta$ assume a discrete set of values between $1$ and $\frac{v\left(t\right)}{c}$.
\item For each discrete value of $\cos \vartheta$, we compute $dF\left(t,\vartheta\right)$ and $t_{\rm a}\left(t,\vartheta\right)$.
\item Then, we change the value of $t$ and repeat,  thus letting $\cos \vartheta$ vary with fixed $t$ over its allowed range, and $t$ vary from the beginning of the emission to the decoupling time, and at each step, we compute all the values of $dF\left(t,\vartheta\right)$ and $t_{\rm a}\left(t,\vartheta\right)$.
\item Now we have a large number of corresponding values of $dF$ and $t_{\rm a}$, and we have to compute $F\left(t_{\rm a}\right)$ as the sum of all the values of $dF$ corresponding to the same $t_{\rm a}$. Unfortunately, we have computed these values numerically, and so they are not exact. Thus some $dF$ that should correspond to the same $t_{\rm a}$ may correspond to slightly different $t_{\rm a}$, and vice versa. So we define $F\left(t_{\rm a}\right)$ as the sum of all the values of $dF$ corresponding to an arrival time between $t_{\rm a}$ and $t_{\rm a}+r_{\rm c}$, where the value of $r_{\rm c}$ is assigned ``by hand." 
\item Of course, during this integration we can see which values of $t$ and $\vartheta$ give us the ``same'' (in the meaning discussed above) value of $t_{\rm a}$, and in this way we obtain plots like the one in Fig. \ref{ETSNCF}.
\end{enumerate}

\begin{figure}
\resizebox{\hsize}{!}{\includegraphics{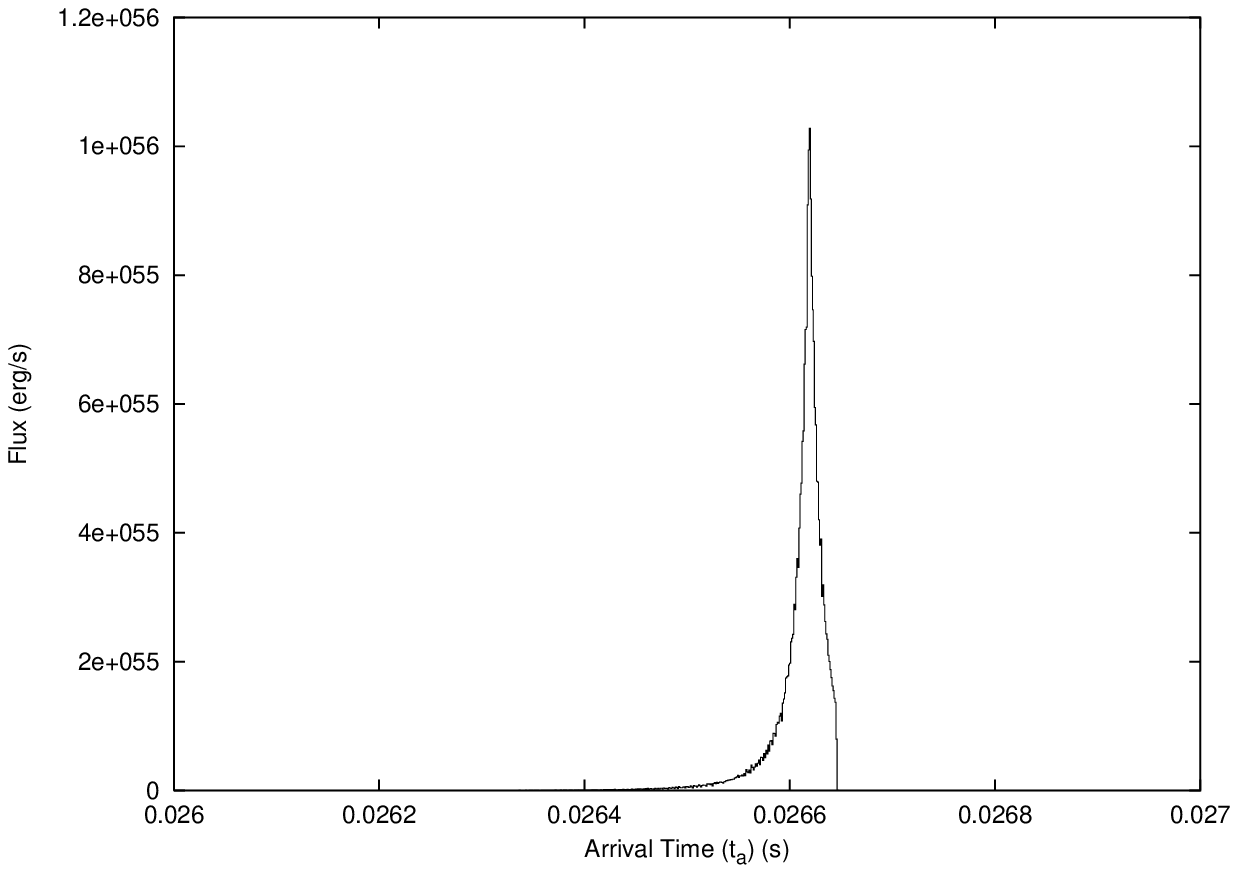}}
\caption{Time profile for a burst from a black hole with $M=10^3M_\odot$ and $Q=0.1Q_{\rm max}$, neglecting fireball thickness effects. The total energy released toward the detector is $E_{\rm tot}\simeq 3.1\cdot 10^{52} {\rm erg}$ and the time duration $\left(T_{\rm 90}\right)$ of the event is $T_{\rm 90}\simeq 7.3\cdot 10^{-5} {\rm s}$. The plot is made with $r_{\rm c}=10^{-6}$ seconds (see text).}
\label{1000_wrong}
\end{figure}

This process yields $F\left(t_{\rm a}\right)$, the so called ``light curve'' of the GRB, and this is shown in Fig. \ref{1000_wrong}, which reveals a very big problem with our computation. In fact, we see that the simulated time profile of the GRB seems to be ``cut'' after the peak, going to zero too fast, while experimentally long ``tails'' are observed after the peak. This is because we have assumed that all photons are emitted from the external surface of the expanding fireball, but they can also be emitted from a certain depth inside the surface, and this implies a time delay in the arrival time for photons emitted from different depths. This effect is almost negligible at the beginning, when the screening factor $S\left(t\right)$ satisfies $S\left(t\right) \ll 1$, but this condition changes very rapidly during the expansion, when $S\left(t\right) \rightarrow 1$. The time required by a photon to cross the entire PEM pulse is about $10^{-2}$ seconds, when $S\left(t\right)\simeq 1$, in the case reported in Fig. \ref{1000_wrong}, while the duration of the peak we have obtained is about $10^{-4}$ seconds. Therefore the thickness effect cannot be neglected and we must change our numerical integration scheme.

\section{The radiation flux with respect to arrival time for a thick fireball}\label{arrivalft}

In the previous section we computed the emitted flux by considering the suppression factor $\sim \frac{\lambda}{D}$ and assuming that all the radiation is emitted from the external surface of the expanding fireball. Now, we relax this last assumption, and modify the estimate of the arrival time given in Eq.(\ref{ta_fin}) by including the delay due to the thickness of the emission region in the fireball. We assume that, at a fixed emission time $t$, only the region of the fireball between $r\left( t \right)$ and $r\left( t \right) - \lambda \left( t \right)$ is active in the photon emission, so the new formula is:

\begin{equation}
t_{\rm a}  = t - \frac{{\int_0^t {v\left( {t'} \right)dt'}  + r_{\rm ds} - r_2 }}{c}\cos \vartheta  + \frac{{r_{\rm ds} }}{c},
\label{ta_fin_thick}
\end{equation}
where $0 \le r_2  \le \lambda \left( t \right)$. This interval, for a fixed value of $t$ and for $\vartheta=0$, leads to a corresponding maximum smearing of the arrival time $t_a$ over the interval

\begin{equation}
\Delta t_a = \frac{\lambda}{c},
\label{intrinsic}
\end{equation}
called the {\em intrinsic time scale}.

Due to this extra factor, the numerical integration is also slightly different:
\begin{enumerate}
\item We pick the first value of $t$ and divide the emitting part of the fireball into $N$ emitting sub-shells, each of width $\frac{\lambda \left( t \right)}{N}$ and characterized by different values of the depth $r_2$, and we assume that the flux emitted by each sub-shell is $\frac{dF}{N}$. We pick the first sub-shell at the surface with $r_2=0$ and vary the value of $\cos \vartheta$ from $\frac{v\left( t \right)}{c}$ to $1$ with a large number of very small finite steps. At each step we compute $dF$ and $t_{\rm a}$. Next we pick the second sub-shell,  increasing $r_2$  by one step, and repeat the computation of $dF$ and $t_{\rm a}$ varying $\cos \vartheta$, and so on for every sub-shell so that the intrinsic time scale of photons traveling within the emitting region is considered.
\item Next we pick the second value of $t$ and divide again the emitting shell of the fireball of width $\lambda \left( t \right)$ into $N$ sub-shells of width $\frac{\lambda \left( t \right)}{N}$. Note that $\frac{\lambda \left( t \right)}{N}$ is an increasing function of $t$, while $N$ is fixed, so these sub-shells are not the same as before. We repeat the computation of $dF$ and $t_{\rm a}$ varying $\cos \vartheta$ for each sub-shell. We pick the third value of $t$ and repeat, and so on until the decoupling time. In other words, we vary $\cos \vartheta$ at fixed $r_2$ and $t$, and vary $r_2$ at fixed $t$.
\item The last task to perform is the same of in previous case, assigning to the arrival time $t_{\rm a}$ the total flux of all the $dF$ corresponding to an arrival time between $t_{\rm a}$ and $t_{\rm a}+r_{\rm c}$.
\end{enumerate}

\begin{figure}
\resizebox{\hsize}{!}{\includegraphics{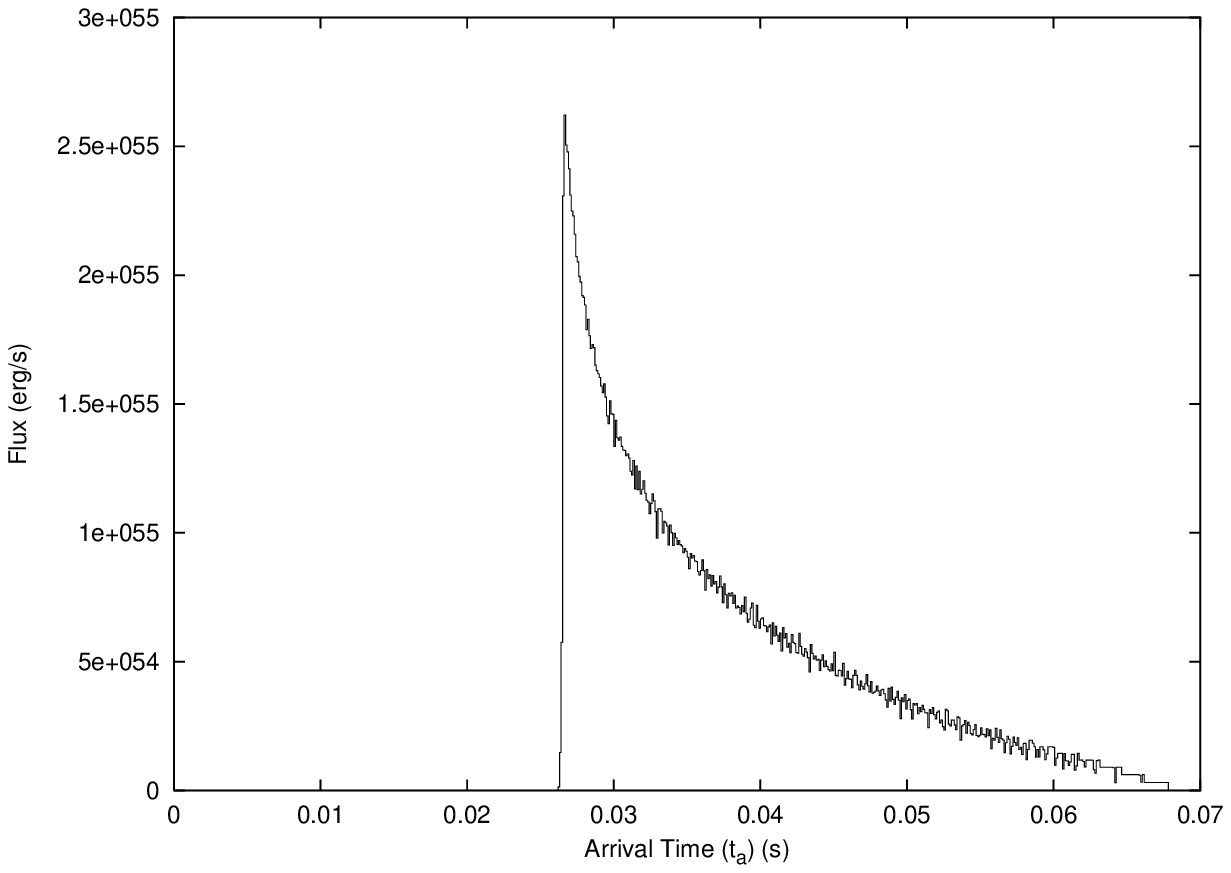}}
\caption{Time profile for a burst from a black hole with $M=10^3M_\odot$ and $Q=0.1Q_{\rm max}$, considering fireball thickness effects. The total energy released towards the detector is $E_{\rm tot}\simeq 2.4\cdot 10^{53} {\rm erg}$ and the time duration $\left(T_{\rm 90}\right)$ of the event is $T_{\rm 90}\simeq 3.0\cdot 10^{-2} {\rm s}$. The plot is made with $r_{\rm c}=10^{-4}$ seconds (see text).}
\label{1000_right}
\end{figure}

\begin{figure}
\resizebox{\hsize}{!}{\includegraphics{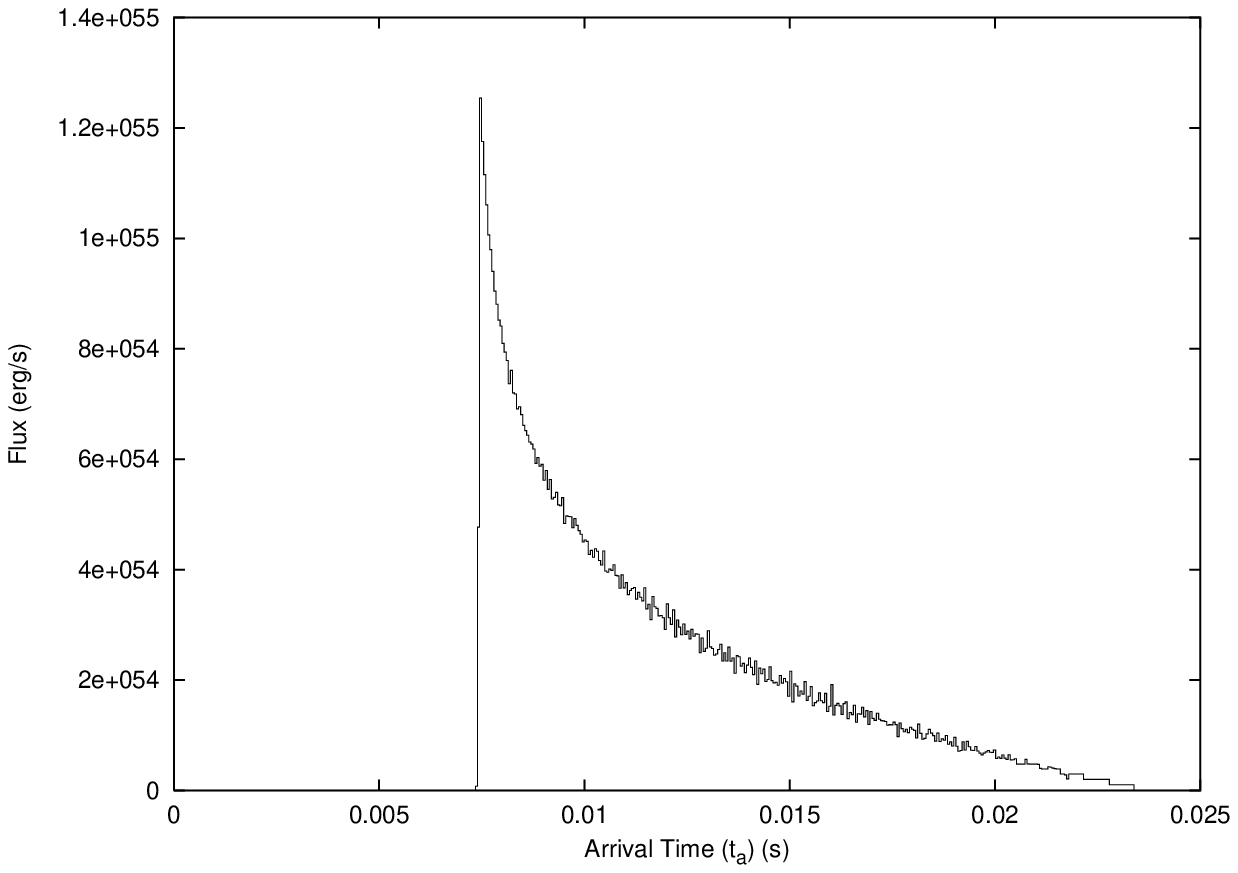}}
\caption{Time profile for a burst from a black hole with $M=10^2M_\odot$ and $Q=0.1Q_{\rm max}$, considering fireball thickness effects. The total energy released toward the detector is $E_{\rm tot}\simeq 4.0\cdot 10^{52} {\rm erg}$ and the time duration $\left(T_{\rm 90}\right)$ of the event is $T_{\rm 90}\simeq 1.1\cdot 10^{-2} {\rm s}$. The plot is made with $r_{\rm c}=5\cdot 10^{-5}$ seconds (see text).}
\label{100_right}
\end{figure}

\begin{figure}
\resizebox{\hsize}{!}{\includegraphics{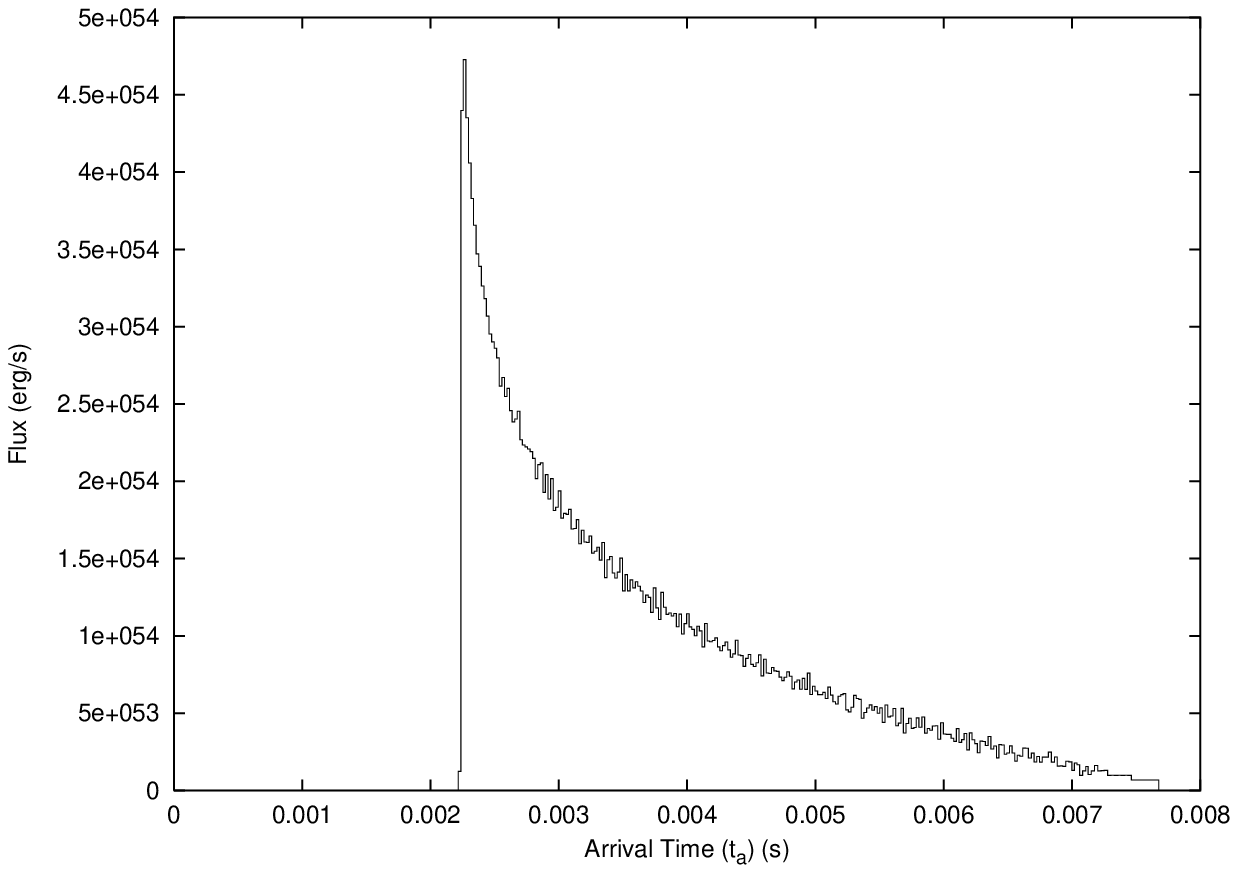}}
\caption{Time profile for a burst from a black hole with $M=10M_\odot$ and $Q=0.1Q_{\rm max}$, considering fireball thickness effects. The total energy released toward the detector is $E_{\rm tot}\simeq 5.2\cdot 10^{51} {\rm erg}$ and the time duration $\left(T_{\rm 90}\right)$ of the event is $T_{\rm 90}\simeq 3.9\cdot 10^{-3} {\rm s}$. The plot is made with $r_{\rm c}=2\cdot 10^{-5}$ seconds (see text).}
\label{10_right}
\end{figure}

\begin{figure}
\resizebox{\hsize}{!}{\includegraphics{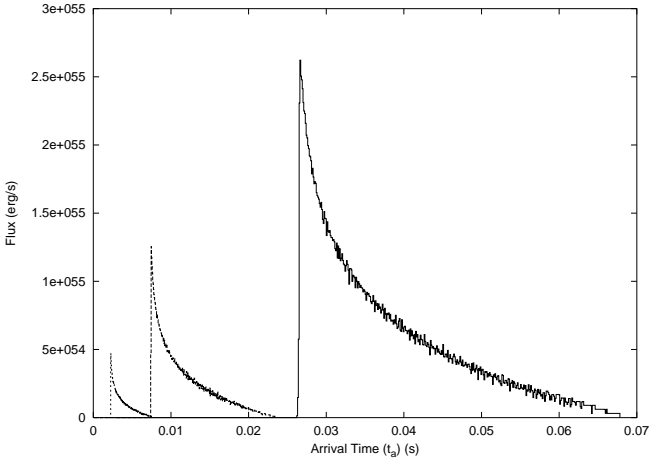}}
\caption{The three peaks of Figs. \ref{1000_right}, \ref{100_right} and \ref{10_right} are shown on the same plot, for visual comparison.}
\label{all_togheter_right}
\end{figure}

The results of this new computation are shown in Figs. \ref{1000_right}, \ref{100_right} and \ref{10_right}, respectively regarding a black hole with mass equal to $10^3M_\odot$, $10^2M_\odot$ and $10M_\odot$ with charge $Q=0.1Q_{\rm max}$, and in Fig. \ref{all_togheter_right} where the three different cases are shown together for comparison purposes.

\begin{figure}
\resizebox{\hsize}{!}{\includegraphics{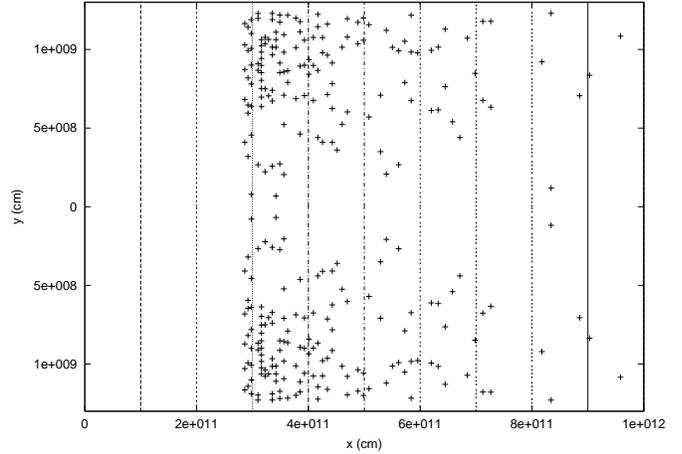}}
\caption{Equitemporal Surfaces for a PEM pulse from a black hole with $M=10^3M_\odot$ and $Q=0.1Q_{\rm max}$, considering fireball thickness effects (see Fig. \ref{1000_right} and par. \ref{arrivalft}), with the same scale as Fig. \ref{ETSNCF}. This plot only shows the points for $t_{\rm a}^\star=0.03$ seconds. The observer is far away along the $x$ axis and the black hole is at the point $\left(0,0\right)$. The vertical lines are a portion of spherical surfaces centered on the black hole. Note that the points does not form definite lines, but show the effect of the superposition of many surfaces emitting independently.}
\label{ETSCONF}
\end{figure}

The new feature here comes from the equitemporal surfaces of the burst. In fact, due to the extra term $r_2$, the relation between $t$ and $\vartheta$ at fixed $t_{\rm a}$ is no longer monotonic. Thus it can happen that photons emitted with a certain $t$, $\vartheta$ and $r_2$ arrive after the ones emitted at a subsequent time $t$, at a larger angle $\vartheta$, but from a lower depth $r_2$. The result is that the points which in Fig. \ref{ETSNCF} lie along curves are now spread on the plane, with no possibility of finding clean surfaces, as can be seen in Fig. \ref{ETSCONF}.

\begin{figure}
\begin{center}
\resizebox{\hsize}{!}{\includegraphics{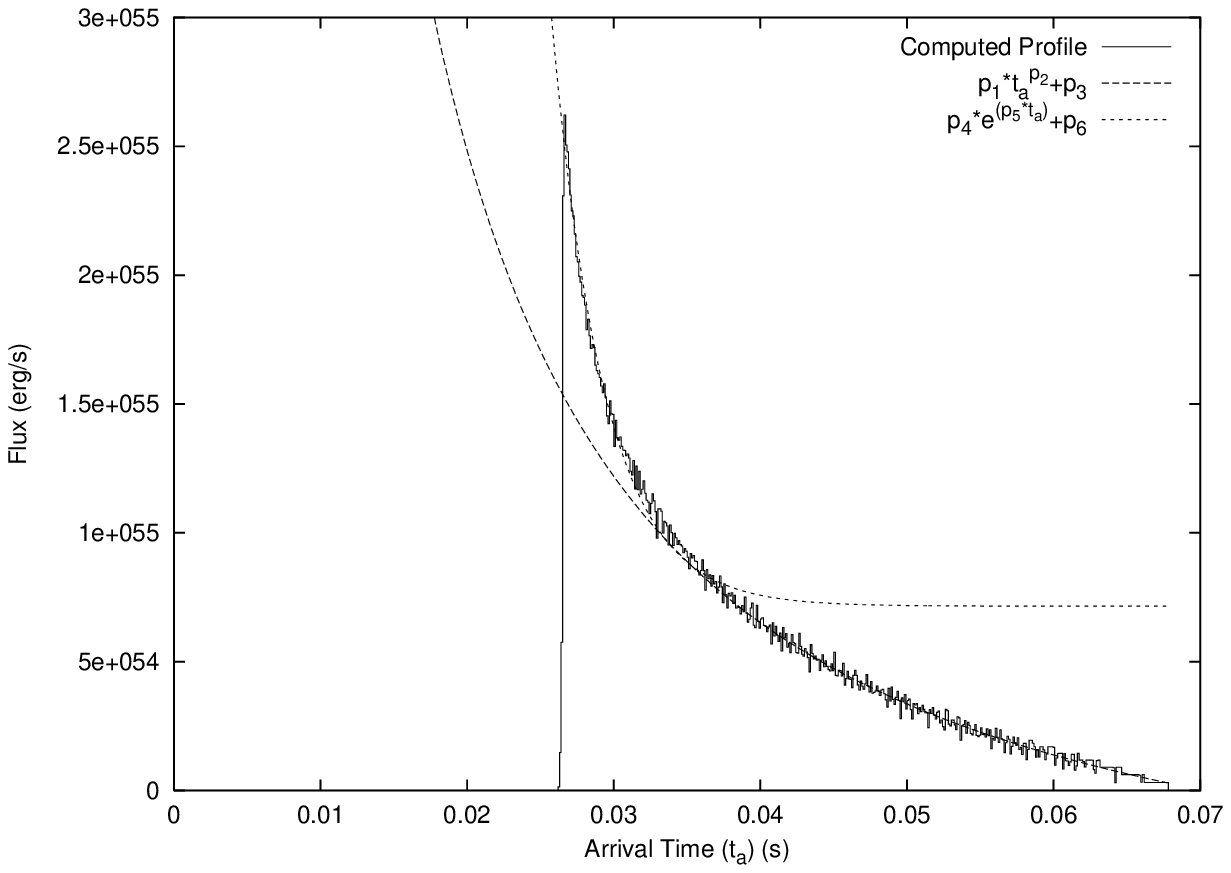}}
\caption{Fit of the ``tail'' of the time profile of Fig. \ref{1000_right}. The values obtained  for the parameters are:\label{1000_fit}}
\begin{tabular}{c|c|c}
\multicolumn{3}{c}{}\\
Parameter & Value & Measure Unit\\
\hline \hline
$p_1$ & $\left( {1,91 \pm 0,85} \right) \cdot 10^{53}$ & $\frac{{\rm erg}}{{\rm s}^{\left(1+p_2\right)}}$\\
$p_2$ & $-1,30 \pm 0,12$ & None\\

$p_3$ & $\left( {-6,02 \pm 0,82} \right) \cdot 10^{54}$ & $\frac{{\rm erg}}{{\rm s}}$\\
\hline
$p_4$ & $\left( {3,09 \pm 0,68} \right) \cdot 10^{58}$ & $\frac{{\rm erg}}{{\rm s}}$\\
$p_5$ & $-280,0 \pm 8,3$ & ${\rm s}^{-1}$\\
$p_6$ & $\left( {7,15 \pm 0,14} \right) \cdot 10^{54}$ & $\frac{{\rm erg}}{{\rm s}}$\\
\multicolumn{3}{c}{}\\
\end{tabular}
\rule{5cm}{0.3pt}
\end{center}
\end{figure}

\begin{figure}
\begin{center}
\resizebox{\hsize}{!}{\includegraphics{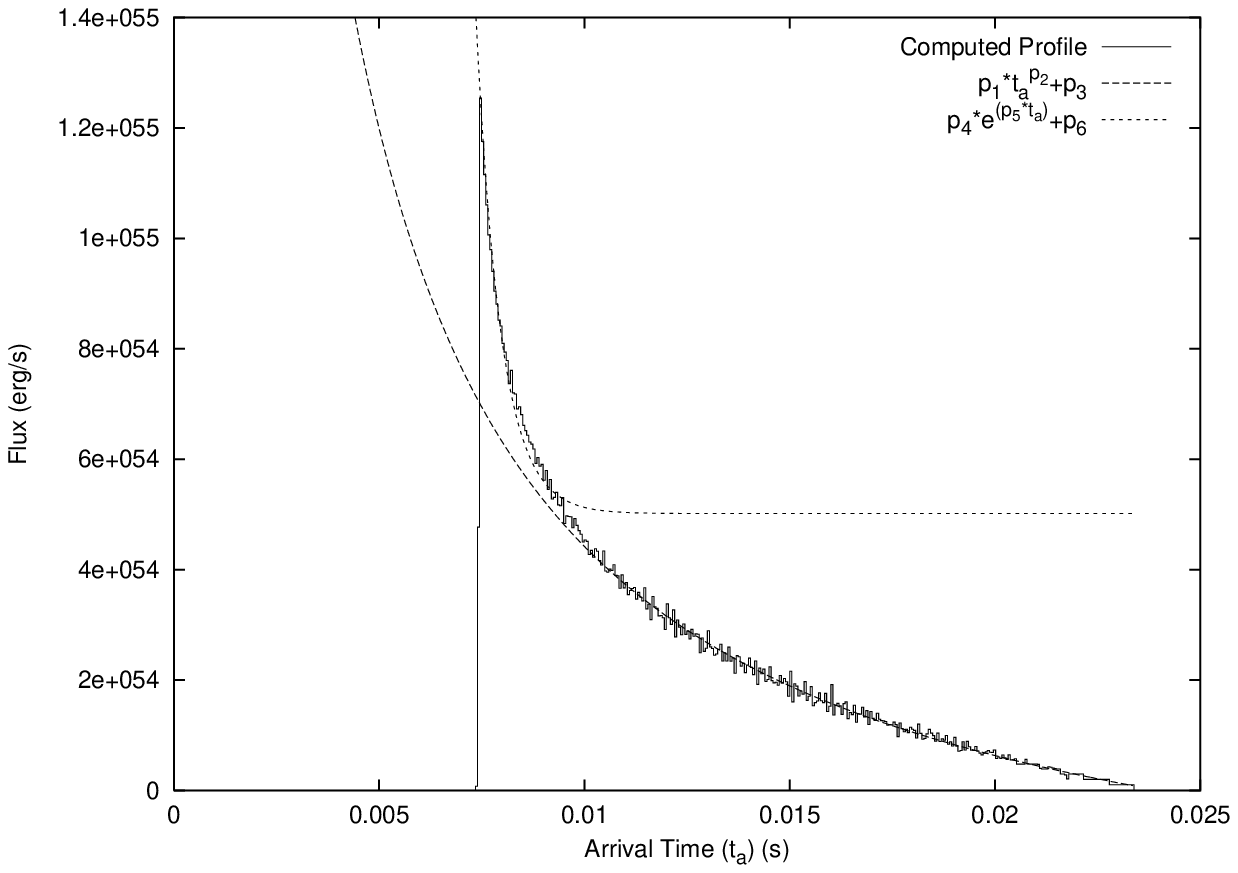}}
\caption{Fit of the ``tail'' of the time profile of Fig. \ref{100_right}. The values obtained  for the parameters are:\label{100_fit}}
\begin{tabular}{c|c|c}
\multicolumn{3}{c}{}\\
Parameter & Value & Measure Unit\\
\hline \hline
$p_1$ & $\left( {7,6 \pm 1,8} \right) \cdot 10^{52}$ & $\frac{{\rm erg}}{{\rm s}^{\left(1+p_2\right)}}$\\
$p_2$ & $-0,999 \pm 0,046$ & None\\
$p_3$ & $\left( {-3,16 \pm 0,23} \right) \cdot 10^{54}$ & $\frac{{\rm erg}}{{\rm s}}$\\
\hline
$p_4$ & $\left( {1,8 \pm 1,1} \right) \cdot 10^{60}$ & $\frac{{\rm erg}}{{\rm s}}$\\
$p_5$ & $-1659 \pm 84$ & ${\rm s}^{-1}$\\
$p_6$ & $\left( {5,015 \pm 0,090} \right) \cdot 10^{54}$ & $\frac{{\rm erg}}{{\rm s}}$\\
\multicolumn{3}{c}{}\\
\end{tabular}
\rule{5cm}{0.3pt}
\end{center}
\end{figure}

\begin{figure}
\begin{center}
\resizebox{\hsize}{!}{\includegraphics{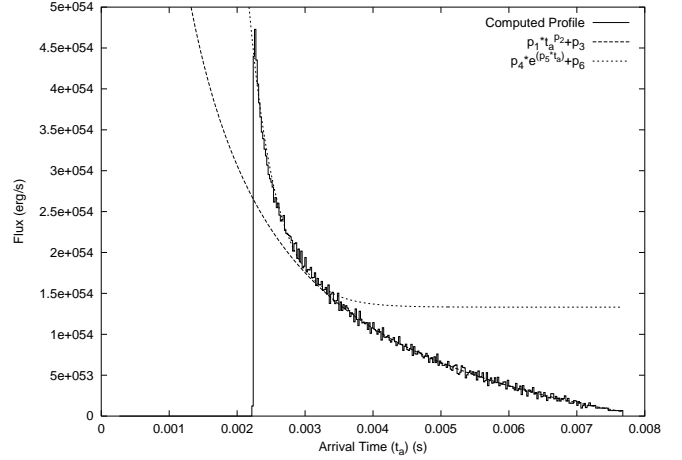}}
\caption{Fit of the ``tail'' of the time profile of Fig. \ref{10_right}. The values obtained  for the parameters are:\label{10_fit}}
\begin{tabular}{c|c|c}
\multicolumn{3}{c}{}\\
Parameter & Value & Measure Unit\\
\hline \hline
$p_1$ & $\left( {1,60 \pm 0,75} \right) \cdot 10^{52}$ & $\frac{{\rm erg}}{{\rm s}^{\left(1+p_2\right)}}$\\
$p_2$ & $-0,900 \pm 0,073$ & None\\
$p_3$ & $\left( {-1,23 \pm 0,15} \right) \cdot 10^{54}$ & $\frac{{\rm erg}}{{\rm s}}$\\
\hline
$p_4$ & $\left( {8,5 \pm 2,5} \right) \cdot 10^{56}$ & $\frac{{\rm erg}}{{\rm s}}$\\
$p_5$ & $\left( {-250 \pm 13} \right) \cdot 10$ & ${\rm s}^{-1}$\\
$p_6$ & $\left( {1,332 \pm 0,048} \right) \cdot 10^{54}$ & $\frac{{\rm erg}}{{\rm s}}$\\
\multicolumn{3}{c}{}\\
\end{tabular}
\rule{5cm}{0.3pt}
\end{center}
\end{figure}

In this case, the light curves are much more similar to the observed ones: we can see that the first part after the peak shows an exponential behavior, followed by a tail with a power law dependence. So we suppose that the second part can be represented by a function like:

\begin{equation}
F = \left( {p_1  \cdot t_{\rm a}^{p_2 } } \right) + p_3 ,
\label{f(x)}
\end{equation}
with $p_1$, $p_2$ and $p_3$ free parameters, and the first part by:

\begin{equation}
F = \left( {p_4  \cdot e^{p_5  \cdot t_{\rm a} } } \right) + p_6 ,
\label{f2(x)}
\end{equation}
with $p_4$, $p_5$ and $p_6$ free parameters. We now apply a ``fit'' algorithm separately on the two parts of the ``tail'' to estimate the parameters. The results are shown in Figs. \ref{1000_fit}, \ref{100_fit} and \ref{10_fit}.

It seems that there is a relation between the values of $p_2$ and $p_5$. In fact, decreasing the value of the black hole mass, and so decreasing the energy of the burst, $p_2$ rises, making the power law decrease slower. In contrast, as $p_5$ decreases the exponential part will be steeper. So the decrease in the energy released in the burst corresponds an ``L'' like shape, with a very steep exponential decrease, followed by an almost constant power-law part.

\section{The temporal structure and time duration}\label{t90}

Within the validity of our approximation, which has the very strong assumption of spherical symmetry, we can now present the main predicted theoretical features of the GRBs which should be compared with observations:
\begin{enumerate}
\item The ``fast-rise, exponential-decay'' profile which represents a clear time trigger of the GRB signal, characterized by an exponential-decay profile with $\sim e^{-\beta t_{\rm a}}$ (see Figs. \ref{1000_fit}, \ref{100_fit} and \ref{10_fit}).
\item The ``tail'' of the total radiation flux with a power law dependence given by $\sim t_{\rm a}^{-\alpha}$, where $\alpha$ is close to unity (see Figs. \ref{1000_fit}, \ref{100_fit} and \ref{10_fit}).
\end{enumerate}
From the temporal profiles of the radiation flux as a function of the EMBH masses we find that the more massive systems have an exponential behavior with a smaller absolute value of the coefficient of the exponent and, correspondingly, a larger absolute value of the exponent of the power law. These effects can be compared and contrasted with GRB observations (see section \ref{arrivalft}).

\begin{figure}
\resizebox{\hsize}{!}{\includegraphics{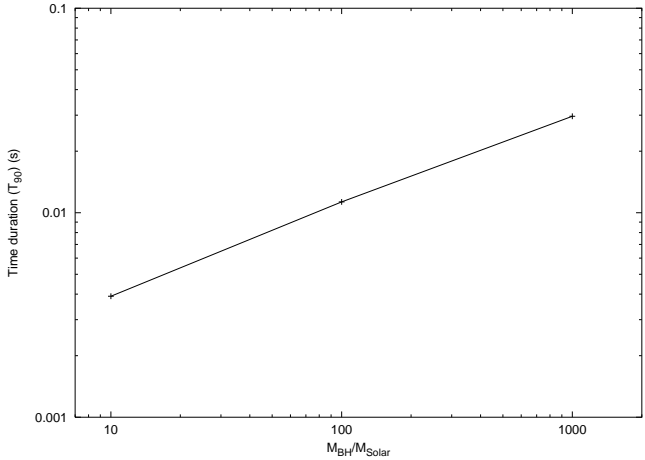}}
\caption{The $T_{90}$ of the three simulated elementary bursts of Fig. \ref{1000_right}, \ref{100_right} and \ref{10_right}, plotted as a function of the black hole mass.}
\label{T90New}
\end{figure}

The time duration of the GRB is usually observationally characterized by the so-called ``$T_{\rm 90}$ criterion'', that is the time interval starting (ending) when the energy detected is the 5\% (95\%) of the total emitted. Namely the $T_{\rm 90}$ is the time duration of the emission of 90\% of the energy. This criterion is very useful in describing observed GRBs, because, due to the background noise, it is very difficult to find the exact starting time and ending time of the emission, while at 5\% and 95\% of the emission the signal is usually well above the noise. We can apply the same procedure to our theoretically computed bursts, obtaining their $T_{\rm 90}$. The results for the three different black holes masses considered are reported in the captions of Figs. \ref{1000_right}, \ref{100_right} and \ref{10_right}. Since now the angular time scale corresponding to different emission epochs are taken into account as well as the intrinsic time scale of the emitting region, the $T_{\rm 90}$ computed are significantly larger (see Fig. \ref{T90New}) than the ones of \cite{rswx99}, computed only on the basis of the angular time scale (see Eq.(\ref{angular})) at the last transparent point.

We emphasize the fact that by breaking the assumption of spherical symmetry that we have adopted and by introducing inhomogeneities in the PEM pulse, the complex sub-structures of the observed GRBs can be easily accommodated in our scenario, using the two different time scales (angular and intrinsic) of the emission process. An inhomogeneity in the PEM pulse with time scale smaller than the angular one (see Eq.(\ref{angular})),

\begin{equation}
\delta < \frac{r}{\gamma^2c\left(1+\frac{v}{c}\right)} \sim O\left(10^{-4}\right),
\label{angmin}
\end{equation}
is invisible in the light-curve. However, any inhomogeneities with time scale $\delta$ within the range between the angular and the intrinsic time scale (see Eq.(\ref{angular}) and Eq.(\ref{intrinsic})),

\begin{equation}
\frac{r}{\gamma^2c\left(1+\frac{v}{c}\right)}<\delta<\frac{\lambda}{c},
\label{angular2}
\end{equation}
would give rise to complex sub-structures in the light curve, up to a number given by

\begin{equation}
\frac{\frac{D}{c}}{\frac{r}{\gamma^2c\left(1+\frac{v}{c}\right)}}.
\label{numbss}
\end{equation}

\section{General considerations about the spectrum}\label{spectraG}

\begin{figure}
\resizebox{\hsize}{!}{\includegraphics{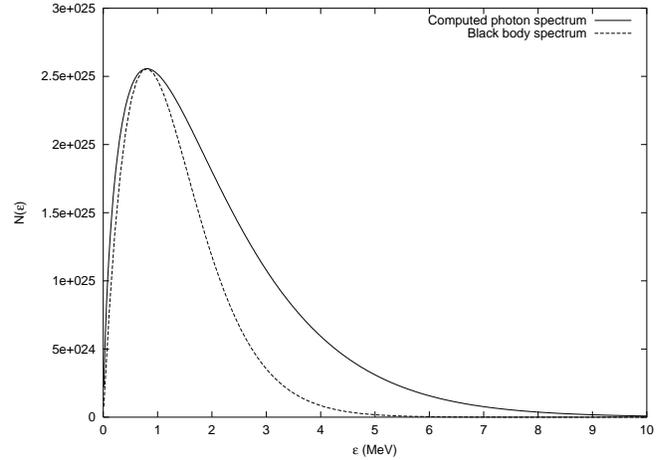}}
\caption{Spectrum of the radiation observed at $t_a=0.03s$ from a black hole with $M=10^3M_\odot$ and $\xi=0.1$, together with a black body spectrum fitted on the peak.}
\label{Spettro1}
\end{figure}

Although the main topic of this article deals with the structure of the burst and its time structure as seen from the observer, we can add some qualitative considerations about the spectrum and especially on its departure from a black body one. If one takes the assumptions of section \ref{flux}, it is easy to compute the expected spectrum of the observed radiation. The observed number spectrum $N_\epsilon$, per photon energy $\epsilon$, per steradian, of photons emitted by a single shell is given by (in photons/eV) (see Eq.(65) of \cite{rswx99}):
\begin{eqnarray}
N_\epsilon(v,T,R) &\equiv&  \int dV {u_\epsilon
 \over \epsilon}= (5.23 \times 10^{11}) 4\pi R^2 dR {\epsilon T \over v
\gamma}\nonumber\\
&\cdot& \log \Biggl[ {1 - exp[- \gamma \epsilon (1 + {v\over c})/T ] \over 1 -
exp[ - \gamma \epsilon (1 - {v\over c})/T ] } \Biggr],
\label{jay:E:nmax}
\end{eqnarray}
which has a maximum at $\epsilon_{max} \cong 1.39 \gamma T\ eV$ for $\gamma \gg 1$. We can then sum this spectrum over an equitemporal surface of our PEM-pulse to get the total spectrum of the radiation observed at a certain arrival time, and this is reported in Fig. \ref{Spettro1}, in the case of a black hole with $M=10^3M_\odot$ and $\xi=0.1$, together with a black body spectrum fitted on the peak, for comparison. It is clear that this spectrum is already different from a black body, both at low and high frequency.

These simplified results may be, however, affected by two additional effects, which are not very relevant for determining the equation of motion of the PEM pulse and the time profile of the bursts, but can indeed be of relevance for the definition of the spectrum. The first modification would affect some of the considerations in section \ref{arrivalft}, where a radial dependence of the temperature of the radiation may be needed following a more detailed description of the transparency condition. This will lead to a further broadening of the spectrum given in Fig. \ref{Spettro1}. The second modification may be due to the consideration of multiple inverse Compton scattering of the photons, described by Kompaneets equation (\cite{K56}; see also \cite{fr72}). Since the increase of energy of a photon per scattering is:

\begin{equation}
\Delta E = \frac{{4kT}}{{m_e c^2 }}E,
\label{DeltaE}
\end{equation}
the condition for the distortion becomes

\begin{equation}
y = \int {\frac{{kT}}{{m_e c^2 }}\sigma_{\gamma,e} n_e dr }=\frac{{kT}}{{m_e c^2 }}\max \left( {\tau_{T} ,\tau _{T}^2 } \right) \ge 1
\label{comptpar}
\end{equation}
with $n_e$ the number of the electrons, $\sigma_{\gamma,e}$ the Thomson cross section and $\tau_T$ the opacity due to Thomson scattering 

\begin{equation}
\tau_T=n_e\sigma_{\gamma,e} D,
\label{taut}
\end{equation}
Then, the number of scatterings is

\begin{equation}
N=\left(\frac{D}{\lambda}\right)^2=\tau_T^2.
\label{nscat}
\end{equation}

In the present model, at emission time, say, $\sim 2$ seconds after the beginning of the expansion, the comptonization appears to be a crucial effect. Indeed, the optical depth for free-free absorption is:

\begin{equation}
\tau _{ff}  = k_{ff} \Sigma \simeq 10^{-8} \ll 1
\label{tauff}
\end{equation}
where

\begin{equation}
k_{ff}  \simeq 0.6 \cdot 10^{23} m_pn_e T^{ - {7 \mathord{\left/
 {\vphantom {7 2}} \right.
 \kern-\nulldelimiterspace} 2}} g^{ - 1} cm^2
\label{kff}
\end{equation}
and $\Sigma$ is the surface density of the electrons. However, the medium is opaque with respect to the Compton scattering (described by the Thomson cross section):

\begin{equation}
\tau_T\simeq 6\cdot 10^4\gg 1
\label{tautn}
\end{equation}
so that the condition for comptonization (Eq.(\ref{comptpar})) is fulfilled.

\begin{figure}
\resizebox{\hsize}{!}{\includegraphics{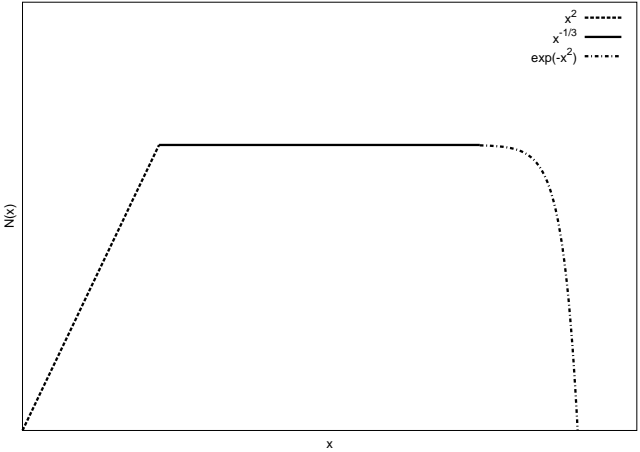}}
\caption{Qualitative plot of the expected spectra due to multiple inverse Compton scattering.}
\label{Spettro2}
\end{figure}

Hence, the black body spectrum, which exists in the comoving frame at the very initial stage of the expansion, quite soon will undergo modifications due to the comptonization. The observed spectrum $F_\nu$, in this case, will depart even more from a thermal one, with  black body distribution at low energies, and a plateau up to the Wien exponential cutoff. In the Fig. \ref{Spettro2} we give a qualitative form of the spectrum.

The variation of the spectrum in the course of the expansion of the fireball can be determined via the simultaneous solution of the hydrodynamical equations of motion and energy balance (given in \cite{rswx99}) together with the Kompaneets equation (\cite{K56}):

\begin{eqnarray}
\frac{{\partial n}}{{\partial y}} = \frac{1}{{x^2 }}\frac{\partial }{{\partial x}}\left[ {x^4 \left( {n + n^2  + \frac{{\partial n}}{{\partial x}}} \right)} \right], \quad x=\frac{h\nu}{kT}\nonumber
\label{KE}
\end{eqnarray}
where $n$ is the occupation number density of photons:

\begin{equation}
n\left(\nu\right)=\frac{c^3}{8\pi h \nu^3}F_\nu
\label{nnu}
\end{equation}

The location of the cutoff, which depends on the number of collisions, will give independent information on the physical parameters of the initial Dyadosphere. This computation is beyond the scope of the present paper.

\section{Conclusions}\label{remarks}

In our model the most general GRB is characterized by six different phases (see e.g. \cite{r00mg9}), each one with distinct physical processes which can lead to specific observational features:

\begin{enumerate}
\item The identification of the precursor. Since the mass range of the EMBH in which the vacuum polarization process can occur varies from $3.2 M_\odot$ to $7.2\cdot 10^6 M_\odot$, it is particularly important to identify the precursors of the collapsing core. While in the case of $\sim 10 M_\odot$ EMBH the understanding of the process of gravitational collapse has been achieved on the ground of binary X-ray sources (see \cite{gr78}), in the case of more massive collapsing systems ($10^5 \sim 10^6 M_\odot$) various scenarios can be considered (see e.g. Gurzadyan \& Ruffini, in preparation). In addition to the identification of the precursor, it is important in our model to identify the magnetohydrodynamical conditions leading to the charge separation process in the collapsing core (see \cite{r00mg9}).  

\item The process of gravitational collapse which leads to the formation of the Dyadosphere. As mentioned in section \ref{maembhm}, the case of spherical symmetry is currently been investigated (\cite{k00mg9}). In nonspherical gravitational collapse, this phase may lead to the observation of gravitational waves and the associated gravitationally induced electromagnetic radiation and electromagnetically induced gravitational radiation (see \cite{jrz73} and \cite{jrz74}).

\item The acceleration of the sole $e^+e^-$ electromagnetic plasma component (the PEM pulse) to relativistic Lorentz $\gamma$ factors greater than $10^2$. The dynamic of this phase has been treated in a previous paper (see \cite{rswx99}). This phase may last all the way to the reaching of the transparency condition. The observational consequence of this phase are the topic of the present paper.

\item The interaction of the PEM pulse, prior to reaching the transparency condition, with some baryonic matter. The possible acceleration to even larger values of the Lorentz $\gamma$ factor ($10^4 \sim 10^5$) of the plasma composed of $\gamma$, $e^+$, $e^-$ and the electron and nucleons of the baryonic matter (the PEMB pulse) has been treated in a previous paper (\cite{rswx00}). The observational aspects of this phase will be presented in a forthcoming paper (Bianco, Ruffini \& Xue, in preparation) where we compute the duration $T_{\rm 90}$ values predicted by this more general model as well as their temporal structure and intensity variation and compare and contrast these results with the observed ones.

\item A substantial part of the energy of the Dyadosphere, in the presence of a large quantity of baryonic matter, will be carried away by the kinetic energy of the baryons. The relative importance of the energy transfer from the Dyadosphere to the kinetic energy of the baryonic component has been estimated in a previous paper (\cite{rswx00}). The observational aspects of this phase, particularly relevant to the observed afterglow and possibly to neutrinos, will be presented in a forthcoming paper (Bianco, Fraschetti, Ruffini \& Xue, in preparation).

\item The acceleration of ultra high energy cosmic rays by the remnant electrodynamic structure of the EMBH (see \cite{ruCNR}). This problem is also under current examination (see Chardonnet \& Ruffini, in preparation).
\end{enumerate}

We have used the detailed model of the creation of photons and electron-positron pairs around an EMBH by the process of vacuum polarization (see \cite{dr75}, \cite{rukyoto} and \cite{prxb}) and their subsequent time evolution (see \cite{rswx99}) in order to compute the different relativistic effects occurring in the determination of the time profile of the observed radiation flux of a GRB with respect to arrival time, as the PEM pulse gradually reaches transparency. If these theoretical predictions will be supported by observational evidence, they will offer an important tool and a strong connection between the observations of GRBs and the properties of the central engine which supplies the energy. This is even more compelling since the simplified model has only two basic parameters, the mass $M$ and charge to mass ratio $\frac{Q}{M}$ of the EMBHs, giving the first clear evidence, in an astrophysical system, of using the extractable mass-energy of a black hole (see \cite{cr71}). In addition, we make precise predictions of the structure of the bursts and of their time variability which can be observationally verified.

The considerations presented in this paper refer only to the above mentioned phase 3 or to the initial evolution of phase 4.  The reaching of transparency by a pure $e^+e^-$ and $\gamma$ component can only occur for very low baryonic matter density around EMBH, $\rho_B \ll 10^{-9} g/cm^3$. This very special circumstance will define a special class of GRB with the specific signature presented in the present paper: short and elementary time variability and absence of afterglow. The estimate of the $T_{90}$ and the shape of the burst may be an important tool for their identification. The spectra features can also be an important tools for verifying the consistency of the model. 

The theoretical tools developed in this paper are applied in the forthcoming publications to the more general case where baryonic matter is present. In this more general case only a small part of the energy of the Dyadosphere will be radiated in the burst. A fraction of the energy, increasing with the amount of baryonic matter, will be transferred to the kinetic energy of the baryonic matter, leading to observational consequences and to the afterglow epoch. 

\appendix

\section{Electron-positron pairs annihilation and GRBs}\label{annihilation}

The evolution of the PEM pulse is completely described by the general relativistic hydrodynamical equations (see \cite{rswx99}) and the rate equation for the number-densities of electrons and positrons 

\begin{eqnarray}
(n_{e^\pm}U^\mu)_{;\mu}&=&(n_{e^\pm}U^t)_{,t}+{1\over r^2}(r^2 n_{e^\pm}U^r)_{,r}\nonumber\\
&=&\bar\sigma \bar v \left[n_{e^-}(T)n_{e^+}(T) - n_{e^-}n_{e^+}\right],
\label{econtin}
\end{eqnarray}
where $\bar\sigma$ is the mean annihilation-creation cross-section of electron-positron pairs, $\bar v$ is the thermal velocity of electron-positron pairs, $n_{e^+}(T)$ and $n_{e^-}(T)$ are the proper number-densities of electrons and positrons, given by appropriate Fermi integrals. We clearly have

\begin{equation}
n_{e^\pm}(T)=n_\gamma(T),
\label{ephoton}
\end{equation}
where $n_\gamma(T)$ is the number-density of photons given by the Bose integral, integrated from $2m_e$ to infinity. 
 
\begin{figure}
\resizebox{\hsize}{!}{\includegraphics{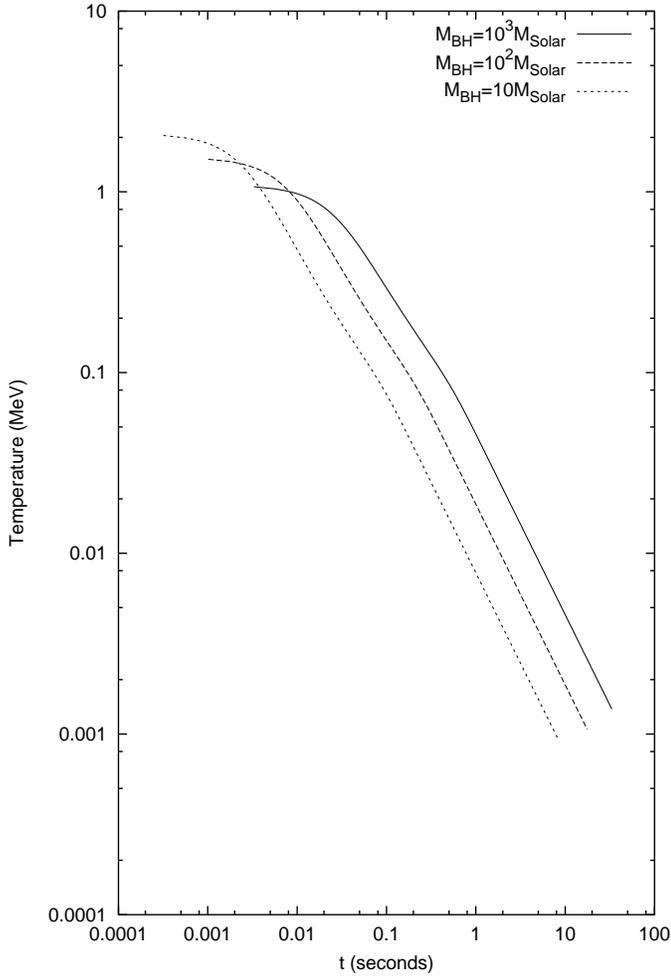}}
\caption{The temperature in the comoving frame as a function of the local laboratory time $t$ 
for typical PEM pulses.
\label{temperature}}
\end{figure}

In the initial evolution stages of the PEM pulses, the temperature $T$ is larger than the energy-threshold $0.5$ MeV of electron-positron pair creation (see Fig. \ref{temperature}). The electrons and positrons created in dyadosphere are in thermal equilibrium with photons, by the process $e^++e^-\leftrightarrow  \gamma +\gamma$.

\begin{equation}
n_{e^\pm} \sim n_{e^\pm}(T)\sim n_\gamma(T),
\label{equi}
\end{equation}
The rate equation becomes
\begin{equation}
(n_{e^\pm}U^\mu)_{;\mu}=0,
\label{econtin0}
\end{equation}
that is just the conservation of the total number of electron-positron pairs (see \cite{rswx99}).

\begin{figure}
\resizebox{\hsize}{!}{\includegraphics{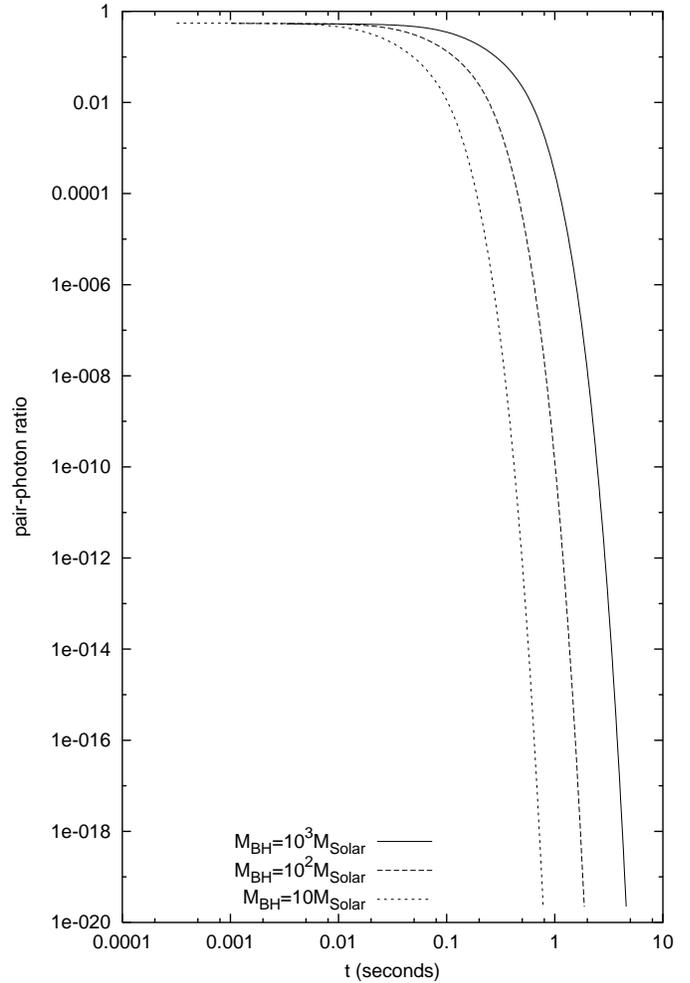}}
\caption{The ratio between number-densities of photons $n_\gamma(T)$ and $n_{\rm e^\pm}$, defined as
$\frac{n_{\rm e^\pm}}{n_\gamma(T)}$, as functions of the local laboratory time $t$ for the three EMBHs of Fig. \ref{screenfig}.}
\label{photon}
\end{figure}

As the temperature $T$ (see Fig. \ref{temperature}) drops first close to and then below the threshold $0.5$ MeV, the electrons and positrons that were in thermal equilibrium with photons annihilate to photons,
\begin{equation}
n_{\gamma}(T)>n_{\rm e^+e^-}
\label{decoupling1}
\end{equation}
when PEM pulses approach transparency. The ratio between the number-densities of photons $n_\gamma(T)$ and pairs $n_{e^\pm}$, defined as
${n_{e^\pm}\over n_\gamma(T)}$, are indicated in Fig. \ref{photon}, for $T\ll 1$MeV, for selected EMBHs.

\section{The screening factor in the radiation flux}

\begin{figure}
\resizebox{\hsize}{!}{\includegraphics{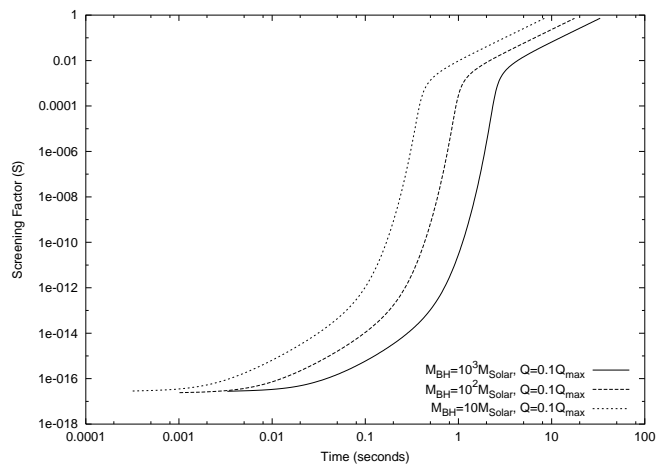}}
\caption{The screening factor $S\left( t \right)$ (see Eq.(\ref{sdit})) for three different EMBHs with charge $Q=0.1Q_{\rm max}$ and mass $M=10M_\odot$, $M=10^2M_\odot$ and $M=10^3M_\odot$ respectively.}
\label{screenfig}
\end{figure}

At the beginning, the PEM pulse is optically thick, so the mean free path $\lambda$ of the photons in the plasma (see Eq.(\ref{Lgamma})) is more than ten orders of magnitude lower than the thickness $D$ of the fireball: for a PEM pulse created by an EMBH with $M=10^3M_\odot$ and $Q=0.1Q_{\rm max}$, the initial numerical values are: $D\simeq 10^9$ cm and $\lambda\simeq 10^{-6}$ cm. So practically we have no radiation emission. During the expansion, the plasma density and temperature both decrease, while $\lambda$ rises, and also the emitted flux rises, until $\lambda\sim D$, when the plasma is optically thin and all the remaining photons escape. We then assume that the black-body radiation flux $F_{\rm BB}$ (in erg/s) emitted by our fireball is given by:

\begin{equation}
F_{\rm BB}\left(t\right) \mathop  \equiv \limits^{def.} S\left( t \right)F_{\rm BB}^{\rm T}\left(t\right)\simeq 4 \pi aT^4r^2c\gamma^2S\left(t\right) ,
\label{Fbb}
\end{equation}
where $a \simeq 1.37 \cdot 10^{26} \frac{\mathrm {erg}}{{\mathrm {cm}}^3{\mathrm {MeV}}^4}$, $T$ is the comoving frame plasma temperature given in ${\mathrm {MeV}}$, $\gamma$ is the Lorentz $\gamma$ factor of the bulk motion of the expanding plasma, $r$ is the radius of the external surface of the PEM pulse and $S\left( t \right)$, is the ``screening factor'', given by (see \cite{prxp99}):

\begin{equation}
S\left( t \right) = \frac{{\sqrt 2 }}{3}\frac{{\lambda  \left( t \right)}}{D}.
\label{sdit}
\end{equation}
In Fig. \ref{screenfig} are plotted the screening factors $S\left(t\right)$ for different plasmas, originated around EMBHs characterized by charge $\frac{Q}{M}=0.1$ and by mass $M=10M_\odot$, $M=10^2M_\odot$ and $M=10^3M_\odot$ respectively.

\begin{figure}
\resizebox{\hsize}{!}{\includegraphics{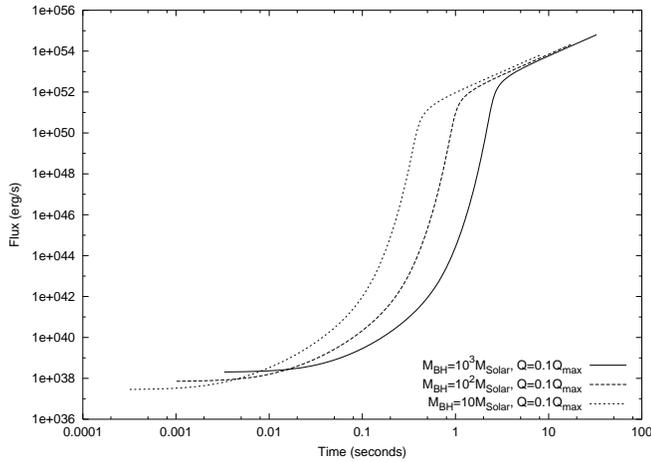}}
\caption{The screened radiation fluxes $F_{\rm BB}\left(t\right)$ (see Eq.(\ref{Fbb})) are shown for the same three cases of Fig. \ref{screenfig}}
\label{fluxfig}
\end{figure}

Using the quantities obtained by the numerical simulation of the expansion of the PEM pulse (see \cite{rswx99}), we can now make a plot of the emitted flux versus emission time $t$, again for selected EMBHs (see Fig. \ref{fluxfig}). A clear radiation flash corresponding to an increase of the order of magnitude of $10^{10}$ in the radiation flux is observed as the screening factor $S\left( t \right)$ approaches unity.

\begin{acknowledgements}

It is a pleasure to thank V.\ Gurzadyan, J.\ Salmonson and J.\ Wilson for helpful discussions, as well as the anonymous referee for attracting our attention on the spectral distributions, complementary to the results of the present work, as well as for the wording of the manuscript.

\end{acknowledgements}


\begin{thebibliography}{}

\bibitem[Christodoulou \& Ruffini, 1971]{cr71}
Christodoulou D., Ruffini R., 1971, Phys. Rev. D 4, 3552

\bibitem[Damour \& Ruffini, 1975]{dr75}
Damour T., Ruffini R., 1975, Phys. Rev. Lett. 35, 463

\bibitem[Djorgovski, Frail, et al., 2000]{FDmg9}
Djorgovski S.G., Frail D., et al., Proceedings of the Ninth Marcel Grossmann Meeting on General Relativity, in press, Rome 2000.

\bibitem[Ehlers, 1971]{eh}
Ehlers J., 1971, Proceedings of Course 47 of the International School of Physics: Enrico Fermi, ed. Sachs, R.K.~, Academic Press, New York.

\bibitem[Felten \& Rees, 1972]{fr72}
Felten J.E., Rees M.J., 1972, A\&A 17, 226

\bibitem[Fenimore, 1999]{F99}
Fenimore E.E., 1999, ApJ 518, 375

\bibitem[Fenimore, et al., 1996]{Fa96}
Fenimore E.E., Madras C.D., Nayakshin S., 1996, ApJ 473, 998

\bibitem[Frontera, et al., 2000]{beppo-sax}
Frontera F., et al., 2000, ApJS, 127, 59

\bibitem[Ghisellini, et al., 2000]{glcr00}
Ghisellini G., Lazzati D., Celotti A., Rees M.J., 2000, MNRAS 316, L45

\bibitem[Giacconi \& Ruffini, 1978]{gr78}
Giacconi R., Ruffini R. (Editors), {\em Physics and Astrophysics of Neutron Stars and Black Holes}, North Holland, Amsterdam 1978

\bibitem[Goodman, 1986]{g86}
Goodman J., 1986, ApJ 308, L47

\bibitem[Granot, et al., 1999]{Ga99}
Granot J., Piran T., Sari R., 1999, ApJ 513, 679

\bibitem[Johnston, et al., 1973]{jrz73}
Johnston M., Ruffini R., Zerilli F., 1973, Phys. Rev. Lett. 31, 1317

\bibitem[Johnston, et al., 1974]{jrz74}
Johnston M., Ruffini R., Zerilli F., 1974, Phys. Lett. 49B, 185

\bibitem[Kompaneets, 1956]{K56}
Kompaneets A.S., 1956, Sov. Phys. JETP 4, 730

\bibitem[Klippert, 2000]{k00mg9}
Klippert R., Miniutti G., Ruffini R., Proceedings of the Ninth Marcel Grossmann Meeting on General Relativity, in press, Rome 2000

\bibitem[M\'{e}sz\'{a}ros \& Rees, 1992]{Mr92}
M\'{e}sz\'{a}ros P., Rees M.J., 1992, MNRAS 258, 41p

\bibitem[Paczy\'{n}ski \& Xu, 1994]{Px94}
Paczy\'{n}ski B., Xu G., 1994, ApJ 427, 708

\bibitem[Panaitescu \& M\'{e}sz\'{a}ros, 1998]{Pa98}
Panaitescu A., M\'{e}sz\'{a}ros P., 1998, ApJ 493, L31

\bibitem[Piran, 1999]{p99}
Piran T., 1999, Phys. Rep. 314, 575

\bibitem[Preparata, et al., 1998a,b]{prxa}
Preparata G.\, Ruffini R.\, Xue S.-S., 1998a, submitted to Phys. Rev. Lett.

\bibitem[Preparata, et al., 1998b]{prxb}
Preparata G.\, Ruffini R.\, Xue S.-S., 1998b, A\&A 338, L87

\bibitem[Preparata, et al., 1999]{prxp99}
Preparata G., Ruffini R., Xue S.-S., in ``Proceedings of the Third ICRA Network Workshop'', Ed. C. Cherubini \& R. Ruffini, S.I.F., Bologna 2000

\bibitem[Ramirez-Ruiz \& Fenimore, 1999]{RRa99}
Ramirez-Ruiz E., Fenimore E.E., 1999, A\&AS 138, 521

\bibitem[Rees, 1966]{r66}
Rees M.J., 1966, Nature 211, 468

\bibitem[Ruffini, 1998]{rukyoto}
Ruffini R., 1998, in \lq\lq Black Holes and High Energy Astrophysics", Proceedings of the 49th Yamada Conference Ed. H. Sato and N. Sugiyama, Universal Ac. Press Tokyo, 1998.

\bibitem[Ruffini, 1999]{ruCNR}
Ruffini R., 1999, A\&AS 138, 513

\bibitem[Ruffini, et al., 1999]{rswx99}
Ruffini R., Salmonson J.D., Wilson J.R., Xue S.-S., 1999, A\&A 350, 334; A\&A Suppl. Ser. 138, 511-512

\bibitem[Ruffini, et al., 2000]{rswx00}
Ruffini R., Salmonson J.D., Wilson J.R., Xue S.-S., 2000, A\&A 359, 855

\bibitem[Ruffini, 2000]{r00mg9}
Ruffini R., Proceedings of the Ninth Marcel Grossmann Meeting on General Relativity, in press, World Scientific, Singapore, 2000.

\bibitem[Ruffini \& Wheeler, 1971]{rw71}
Ruffini R., Wheeler J.A., 1971, Phys. Tod. 24 (January), 30

\bibitem[Sari \& Esin, 2000]{SE00}
Sari R., Esin A.A., 2000, submitted to ApJ, astro-ph/0005253

\bibitem[Sari \& Piran, 1997]{Sa97}
Sari R., Piran T., 1997, ApJ 485, 270

\bibitem[Sumner \& Fenimore, 1997]{Sua97}
Sumner M.C., Fenimore E.E., 1997, in {\em Gamma Ray Bursts, Proceedings of the $4^{th}$ Huntsville Symposium}, eds. Meegan, Preece, Koshut

\bibitem[Vietri, 1998]{V98}
Vietri M., 1998, Phys. Rev. Lett. 80, 3690

\end{thebibliography}
\end{document}